\documentclass[openacc]{rstransa}

\usepackage{caption} 
\usepackage{subcaption} 

\newcommand{\V}{\mathrm{Var}}

\newcommand{\m}[1]{\mathrm{#1} }
\renewcommand{\cal}[1]{\mathcal{#1}}
\renewcommand{\v}[1]{\boldsymbol{#1}}
\newcommand{\bb}[1]{\mathbb{#1}}

\begin{document}

\title{Being Bayesian in the 2020s: opportunities and challenges in the practice of modern applied Bayesian statistics}

\author{
Joshua J. Bon$^{1,2}$, Adam Bretherton$^{1,2}$, Katie Buchhorn$^{1,2}$, Susanna Cramb $^{1,4}$, Christopher Drovandi$^{1,2}$, Conor Hassan$^{1,2}$, Adrianne L. Jenner$^{1,2}$, Helen J. Mayfield $^{1,5}$, James M. McGree$^{1,2}$, Kerrie Mengersen $^{1,2}$, Aiden Price $^{1,2}$, Robert Salomone$^{1,3, \star}$, Edgar Santos-Fernandez $^{1,2}$, Julie Vercelloni $^{1,2}$, Xiaoyu Wang$^{1,2}$ \\ }

\address{$^{1}$ Centre for Data Science, Queensland University of Technology\\
$^{2}$ School of Mathematical Sciences, Queensland University of Technology \\
$^{3}$ School of Computer Science, Queensland University of Technology \\
$^{4}$ School of Public Health and Social Work, Queensland University of Technology\\ 
$^{5}$ School of Public Health, The University of Queensland \\ 
$^\star$ robert.salomone@qut.edu.au
}



\begin{abstract}
Building on a strong foundation of philosophy, theory, methods and computation over the past three decades, Bayesian approaches are now an integral part of the toolkit for most statisticians and data scientists. Whether they are dedicated Bayesians or opportunistic users, applied professionals can now reap many of the benefits afforded by the Bayesian paradigm. 
In this paper, we touch on six modern opportunities and challenges in applied Bayesian statistics: intelligent data collection, new data sources, federated analysis, inference for implicit models, model transfer and purposeful software products. 
\end{abstract}


\begin{fmtext}

\end{fmtext}


\maketitle

\section{Introduction}

Bayesian data analysis is now an established part of the lexicon in contemporary applied statistics and machine learning. 
There is now a wealth of practical know-how to complement the continued development and increasing access to Bayesian models, algorithms, and software.
There is also a weighty body of published case studies that testify to the successful implementation and associated benefits of the Bayesian paradigm in practice. 
However, as with all fields of knowledge, the task is unfinished: each success begets further opportunities and challenges, which in turn drive new directions for innovation in research and practice. 
In this paper, we identify six such directions that, among many others, are driving the evolution of applied Bayesian modelling in this decade. 
For each of these, we provide a brief overview of the issue and a case study that outlines our experience in practice. 

The first direction focuses on intelligent data collection: instead of collecting and analysing all possible data, or alternatively relying on traditional static experimental or survey designs, can we devise efficient, cost-effective approaches to collecting those data that will be most informative for the inferential purpose? 
In Section 2, authors Buchhorn and McGree focus on the opportunity to address this issue through Bayesian optimal experimental design. While there is an emerging literature on this approach in the context of clinical trials, they extend this attention to sampling designs for complex ecosystems. 
Furthermore, they address the challenge of exact implementation of the derived design in practice by introducing sampling windows in the optimal design. The new methodology and computational solution are illustrated in a case study of monitoring coral reefs. 

Following from consideration of data collection, the second direction considered in this paper focuses on opportunities and challenges afforded through the emergence of new data sources. 
In Section 3, authors Price, Santos-Fern{\'a}ndez, and Vercelloni focus on two such sources: quantitative information elicited from subjects in virtual reality settings, and data provided by citizen scientists. 
Bayesian approaches to modelling and analysing these data can help to increase trust in these data and facilitate their inclusion in mainstream analyses. 
Some methods for achieving this are set in the context of two case studies based in the Antarctic and the Australian Great Barrier Reef.

The challenges of data collection are considered from a different direction in Section 4. 
Here, authors Hassan and Salomone reflect on the exponential rise in interest in federated analysis and learning. 
A canonical application of these approaches is the analysis of sensitive data from multiple data sources held by different data custodians, while leaving the data in situ and maintaining data privacy.  
The case study in this section focuses on federated learning with spatially-dependent latent variables.

In Sections 5 and 6 we swing attention away from data to the models themselves. 
First, authors Drovandi, Jenner, Salomone, and Wang consider the challenge of modelling increasingly complex systems via implicit models, i.e., models with intractable likelihoods that can nevertheless be simulated, and the opportunity afforded by likelihood-free algorithms such as Sequential Monte Carlo based Approximate Bayesian Computation (SMC-ABC). 
These approaches are applied to a substantive case study of calibrating a complex agent-based model of tumour growth. 
In Section 6, another direction for modelling is discussed by authors Bon, Bretherton and Drovandi. This focuses on the challenge of transferring models developed in one context (dataset, location, etc.) to another context. 
Fully Bayesian approaches to this challenge are still emerging and promise great opportunities in both research and practice. 

The final direction we explore is in the translation of Bayesian practice to software products. 
We acknowledge the plethora of Bayesian packages embedded in software such as \verb|R|, \verb|Matlab|, and \verb|Python|, as well as stand-alone Bayesian products such as \verb|BUGS|, \verb|INLA|, and \verb|Stan|. 
These have revolutionised the practice of Bayesian data analysis and have placed this capability in the hands of applied researchers and practitioners. 
In Section 7, we focus on substantive software products created to support purposeful decision-making that are underpinned by Bayesian models. Author Mayfield describes a COVID-19 vaccine risk-benefit calculator (CoRiCAL) driven by a Bayesian network model; Vercelloni describes a platform for global monitoring of coral reefs (ReefCloud) based on a Bayesian hierarchical model; and Cramb describes an interactive visualisation of small area cancer incidence and survival across Australia (the Australian Cancer Atlas) based on a Bayesian spatial model. 

\section{Intelligent Data Collection}
\subsection{Overview}
The ability to determine and characterise underlying mechanisms in complex systems is paramount to pioneering research and scientific advancement in the modern era.
Over the last decade, the rise of data generation from sensor and internet enabled devices has catalysed the advancement of data collection technologies and analysis methods used to extract meaningful information from complex systems.
However, the sheer size of these complex systems (e.g., natural ecosystems like the Great Barrier Reef and river networks) and the expense of data collection means that data cannot be collected throughout the whole system.
Further, practical constraints like connectivity, accessibility, and data storage issues reduce our ability to sample frequently through time.
This has led to innovation in statistical methods for data collection, promoting an emerging era of ``intelligent data collection'' where data are collected for a particular purpose such as understanding mechanisms for change, monitoring biodiversity, and identifying threats or vulnerabilities to long-term sustainability.
Bayesian optimal experimental design is one such area of recent innovation.

Bayesian design offers a framework for optimising the collection of data specified by a design~$\v d$ for a particular experimental goal, which may be to increase precision of parameter estimates, maximise prediction accuracy and/or distinguish between competing models.
More specifically, Bayesian design is concerned with maximising an expected utility, $U(\v d) = \mathbb{E} \, u(\v d, \boldsymbol{\theta}, \v y) $ through the choice of design $\v d$ within a design space $\mathcal{D}$, while accounting for uncertainty about, for example, the parameter $\boldsymbol{\theta} \in \Theta$ and all conceivable data sets we might observe $\v y \in \mathcal{Y}$. 
A Bayesian optimal design $\v d^*$ can therefore be expressed as,
\begin{equation}\label{optimal_d}
\v d^* = \mbox{arg max}_{\v d \in \mathcal{D}} \mathbb{E} \, u(\v d, \boldsymbol{\theta}, \v y) \nonumber = \mbox{arg max}_{\v d \in \mathcal{D}} \int_{\mathcal{Y}} \int_{\Theta} u(\v d, \boldsymbol{\theta}, \v y) p(\boldsymbol{\theta}, \v y; \v d)  d\boldsymbol{\theta} d\v y,  
\end{equation}
where $p(\boldsymbol{\theta}, \v y ;  \v d)$ defines the joint distribution of $\boldsymbol{\theta}$ and $\v y$ given a design $\v d$.


Unfortunately, determining $\v d^*$ can be challenging.
Firstly, the utility function $u(\v d, \boldsymbol{\theta}, \v y)$ typically involves computing some form of expected value with respect to the posterior distribution, which itself is typically analytically intractable. Further, $U(\v d)$ is itself an expectation taken with respect to the prior-predictive distribution, which is also typically intractable. This means numerical or approximate methods are needed, which may impose substantial compute time and/or require a stochastic approximation.
For example, Monte Carlo integration has been proposed as an approach to form an approximation to the expected utility as follows:
\begin{equation}
    \v d^* \approx \mbox{arg max}_{\v d \in \mathcal{D}} \frac {1}{M} \sum^{M}_{m=1} u(\v d, \boldsymbol{\theta}^{(m)}, \v y^{(m)}), \label{montecarlo_d} 
\end{equation}
where $\boldsymbol{\theta}^{(m)} \sim p(\boldsymbol{\theta};\v d)$ and $\v y^{(m)} \sim p(\v y | \boldsymbol{\theta}^{(m)}; \v d)$, for some large value of $M$.  Thus, computations involving $M$ different individual posterior distributions are required just to approximate the expected utility of a design.

Secondly, $\v d$ may be high-dimensional, meaning that a potentially large optimisation problem needs to be solved for a computationally expensive and noisy objective function.
Accordingly, the majority of research in Bayesian design has focused on developing new methods to address one of these two challenges.
Below we provide a brief summary of some of the relevant literature from the last 25 years.

Since the conception of statistical decision theory \cite{neyman1928use, wald1949statistical} upon which the decision-theoretic framework of Bayesian design is based \cite{lindley1972bayesian}, there have been numerous strategies presented in the literature to address the above challenges. 
Curve-fitting methods were proposed by \cite{muller1995optimal} to the approximate expected utility in Equation \eqref{montecarlo_d}.
Here, the fitted curve is optimised as a surrogate for the true expected utility to determine the choice of (approximately optimal) design.
An alternative simulation-based method proposed by \cite{muller1999simulation} formed the following augmented joint distribution on $\v d$, $\boldsymbol{\theta}$, and $\v y$: 
\begin{equation*}
    h_J(\v d, \boldsymbol{\theta}_{1:J}, \v y_{1:J})\propto\prod_{j=1}^{J}u(\v d, \boldsymbol{\theta}_{j}, \v y_{j})p(\v y_{j}, \boldsymbol{\theta}_{j}; \v d),
\end{equation*}
where it can be shown that the marginal distribution of $\v d$ is proportional to $U(\v d)$.  Markov chain Monte Carlo (MCMC) methods were then used to sample from this distribution, and subsequently to approximate the marginal mode of $\v d$.
Extensions of this approach were given by \cite{muller2004optimal, amzal2006bayesian} which include adopting a sequential Monte Carlo algorithm to more efficiently sample from the augmented distribution as $J$ increases.
However, such approaches are limited to low-dimensional design problems (i.e., \ 3--4 design points) and simple models due to difficulties in sampling efficiently in high dimensions.
 
Recently, there has been a shift from sampling-based methods to rapid, approximate posterior inference methods. 
Combined with a Monte Carlo approximation as given in Equation \eqref{montecarlo_d}, this has enabled expected utility functions to be efficiently approximated for realistic design problems.
This includes those based on complex models (such as non-linear models) and models for data that exhibit complex dependence structures (such as those with different sources of variability including spatially and between groups).
Such approximate inference methods include the Laplace approximation \cite{overstall_normal} and variational Bayes \cite{foster_2019}, which have been combined with new optimisation algorithms (e.g., the approximate coordinate exchange algorithm; ACE \cite{ace}) to solve the most complex and high-dimensional design problems to date.

The most prominent application of Bayesian design methods appears in the clinical trial literature \cite{berry2006bayesian}.
Recently, this has been exacerbated by the outbreak of COVID-19 where it has been desirable to conduct clinical trial assessments as quickly as possible, with Bayesian (adaptive) designs shown to yield more resource efficient and ethical clinical trials \cite{connor_2013,thorland_2018}.
More recently, Bayesian design methods have been proposed as a basis to efficiently monitor large environmental systems like the Great Barrier Reef \cite{kang_2016,thilan_2022}.
In the following case study, we show how such methods can be used to form sampling designs to monitor a coral reef system, and extend these methods to provide flexible designs that address major practical constraints when sampling real-world ecosystems.
\subsection{Case Study: Sampling Windows for Coral Reef Monitoring}
Coral reefs are biodiversity hot-spots for marine species under threat from anthropogenic impacts related to climate change, water pollution and over-exploitation, among other factors \cite{wagner2020coral}.
Coral cover is a commonly used indicator to infer the health of coral reef environments \cite{ltmp2022}, where data collection relies on a series of images taken underwater, along a transect (a line across a habitat).
Monitoring of coral reef environments is expensive in terms of monetary, human and technological costs, particularly for remote locations.
Informative data is critical to support conservation decisions, but with limited resources to invest in monitoring programs, the need is to optimise in-field activities that will result in the intelligent collection of data.
Following \cite{buchhorn2022bayesian}, we consider monitoring submerged shoals which are coral reefs that exist at depths of around 18 to 40m below sea level.  
Data collection at such depths requires unmanned vehicles to be deployed along a design, i.e., a series of transects which specify where images should be collected.  
However, spatially precise sampling is known to be difficult in deeper reefs due to unpredictable weather and water currents.
Therefore, our aim is to provide Bayesian designs that offer flexibility in where transects will be placed while taking into consideration the complex nature of the systems we are monitoring such as the spatial dependence of natural processes.

In order to define a design, we specify the placement of each transect $k=1,\ldots, q$ on the shoal by its midpoint given in Easting and Northing coordinates, i.e., $E_k$ and $N_k$, and the angle of the transect, $\alpha_k$, in degrees.
Each transect line is expressed as a design point $\v d_k = (E_k, N_k, \alpha_k)$.
The exact sampling locations (equally spaced along the fixed-length transect) are specified as $\v s_i$.
For each transect, we introduce a radius parameter $r_k > 0$ for the purpose of allowing the sampled image locations to disperse by $\delta_{1}, \delta_{2} \stackrel{\rm iid}\sim \mbox{Unif}(-r_k, r_k)$, i.e., sampling at $\v s + \v \delta$.
For image $i$, a number, $n_i$, of randomly selected points on the image are classified as either hard coral or not.
Accordingly, the number of points within an image that contain
hard coral, $y_i$, is modelled as
\begin{align*}
    y_i|\boldsymbol{\beta},  Z_i &\sim \text{Binomial}\left(n_i, \text{logit}^{-1} (\v \beta^\top \v x_i + Z_i)\right),\\
    \textbf{Z} &\sim \mathcal{N}(\v 0, \Sigma(\boldsymbol{\gamma})),
\end{align*}
for regression parameters $\boldsymbol{\beta} = (\beta_0,...,\beta_n)^\top$, covariance kernel parameters $\boldsymbol{\gamma} = (\gamma_1,...,\gamma_m)^\top$ for spatially-correlated random effect $\textbf{Z}$, and  covariates $\v x_i$.
The priors for $\beta$ and $\gamma$ are based on consideration of historical data (depth and depth squared) collected on the shoal. See \cite{buchhorn2022bayesian} for further details.

As a basis for improved monitoring, we consider the amount learned from the data regarding parameters of the above model as our goal of data collection. 
For this, we specify our utility function as the Kullback-Liebler divergence of the posterior from the prior distribution, where larger values suggest the data is more informative with respect to model parameters.

\begin{figure}
\includegraphics[width=\textwidth]{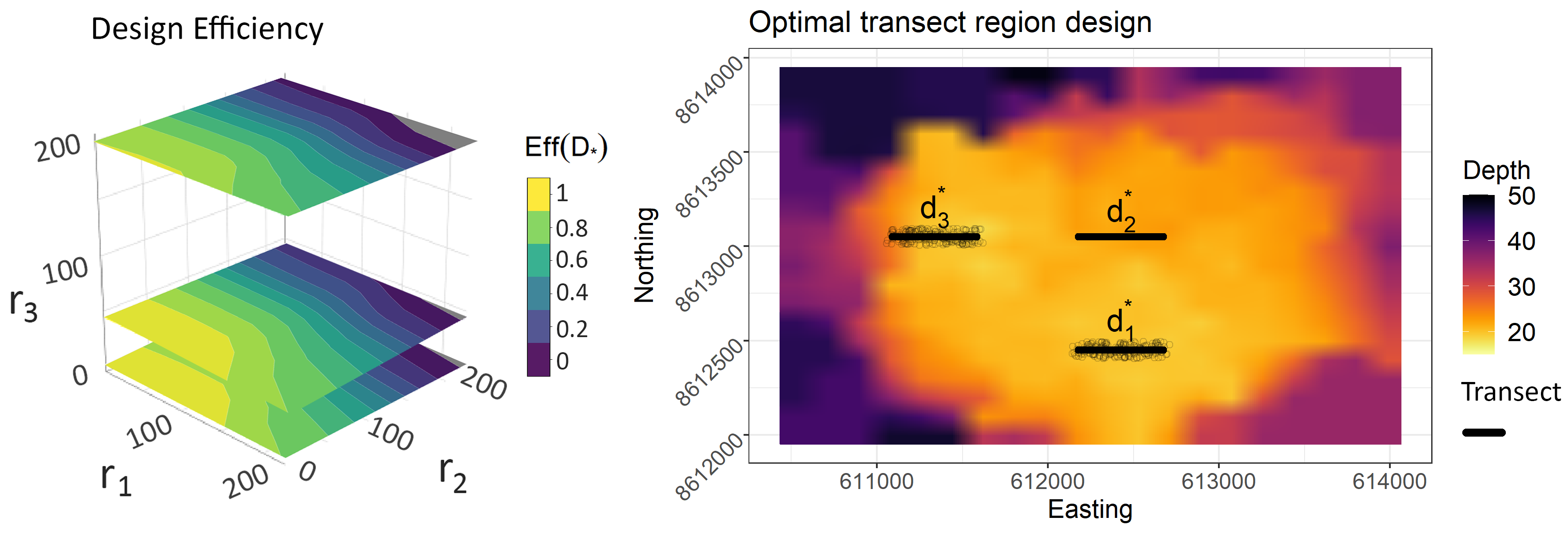}
\caption{The Bayesian design across the Barracouta East coral shoal, $\v d^* = (\v d_1^*, \v d_2^*, \v d_3^*)$, are illustrated as black transect lines (right). 
Sampling windows are formed around these transects allowing for flexibility in sampling locations while retaining 0.99 of the optimal utility.
Design efficiency contours across $\v r \in \bb R^q$ are shown (left).
}\label{figutility} 

\end{figure}

For a computationally-efficient approximation of the utility for a given design, we employ a Laplace approximation of the posterior distribution, i.e., an approximation of the form 
\begin{equation*}
    \mathcal{N}\bigg(\boldsymbol{\theta}^{*},\m H(\boldsymbol{\theta}^{*})^{-1}\bigg),
\end{equation*}
where $\boldsymbol{\theta}^* = \mbox{arg max}_{\boldsymbol{\theta} \in \Theta} \mbox{log } p(\v y, \boldsymbol{\theta}; \v d)$ and $\m H(\boldsymbol{\theta}^{*})$ is the Hessian matrix evaluated at $\v \theta^*$. Here $\v \theta = (\v \beta, \v \gamma)$, and marginalization of $\v Z$ is performed approximately using Monte Carlo integration. To obtain an optimal design to for monitoring of the shoal, we propose a two-step approach: \begin{enumerate} 
\item Firstly, a global search for the Bayesian optimal design $\v d^* = (\v d_1^*, ..., \v d_q^*)$, where $q=3$ (the total number of transects) is conducted. We consider a discretised design space, and find designs via a discrete version of ACE; and 
\item  Secondly, we form design efficiency windows (illustrating robustness to imprecise sampling) across $r_k$ for each transect $k=1,\ldots, q$.
To do so, we specify a zero-mean Gaussian process (GP) prior for the approximate expected utility across $\v r \in \bb R^{q}$ by $\widehat{U}(\v r ; \v d^*)$, i.e., \ $\widehat{U}(\v r ; \v d^*) \sim \mathcal{GP}(\textbf{0}, \m K(\cdot) + \zeta_0 \m I)$, for some kernel matrix $\m K(\cdot)$, and $\zeta_0 > 0$. 
The windows are then obtained as follows:
\begin{enumerate}
\item Center the radius on $\v d^*$, i.e.,\ the Bayesian design from (i), and specify a maximum value for $r_k$ for $k=1,\ldots,q$;

\item Randomly sample $\delta_{i,1}, \delta_{i,2} \stackrel{\rm iid}{\sim} \mbox{Unif}(-r_k, r_k)$, where $k$ is the transect from which image $i$ is obtained,  and evaluate the approximate expected utility of the design at locations $\v s_i + \v \delta_i$;
\item Fit a Gaussian process defined on $\v r \in \bb R^q$ to the approximate expected utilities;
\item Emulate the expected utility surface across values of $\v r$ using the posterior predictive mean of the Gaussian process, denoted $\bar{U}(\v r)$; 
\item Normalise the predicted expected utility values by that of the original Bayesian design as follows:
\begin{equation}
{\rm eff}(\v r) = \frac{\bar{U}(\v r ; \v d^*)}{\bar{U}(\v 0 ; \v d^*)},
\label{efficiency_contours}
\end{equation}
and use the above to obtain design efficiency contours (plotted in Figure \ref{figutility}, left). For some design efficiency contour value $c > 0$, the corresponding sampling window is the region in space defined by radii $\v r(c)$ that satisfy ${\rm eff}(\v r(c)) = c$.
\end{enumerate}
\end{enumerate}

Based on the approach, the Bayesian design, $\v d^*$, shown in Figure \ref{figutility}, situates the transects in shallower areas of the reef, but at different depths in the shallow areas, presumably to provide information about the depth effects, $\boldsymbol{\beta}$.
Avoiding the deeper regions of the shoal makes sense physiologically, as the corals monitored here are photosynthetic organisms, and therefore rely on light to survive.
This design thus avoids the collection of data in areas where there is little chance of observing coral.
The design efficiency contours are also shown in Figure \ref{figutility}.
If we consider a design efficiency of 0.99 in Equation \eqref{efficiency_contours}, then possible radius values are $(50, 0, 44)$ for the three transects, with the flexibility (sampling windows) this provides shown around each transect.  
As can be seen, transect $\v d^*_2$ is more sensitive than the other two, suggesting more effort should be placed in sampling this transect precisely.
In practical terms, sampling from shallow areas of the reef, $\v d^*_1$ and $\v d^*_3$, can be undertaken when the conditions are more unpredictable (e.g., strong currents), and samples from $\v d^*_2$ can be obtained when field conditions are more preferable.

In conclusion, Bayesian optimal design addresses a fundamental problem in science: the intelligent collection of data resulting in greater information efficiency, reduced sampling cost and improved estimation. 
Such benefits have been observed in clinical trials \cite{bassi2021bayesian, giovagnoli2021bayesian, kojima2021early,mcgree2022controlled} and environmental monitoring \cite{leach2022recursive, thilan_2022}, and we have shown how they can be used to offer flexible yet efficient sampling in a real-world context.
One limitation of the approach is the potential reliance of designs on a number of assumptions, e.g., an assumed model for the data, so we would encourage future research in areas that reduce this reliance and thus provide more robust designs for data collection.

\section{New data sources}

As part of the digital revolution, data from new types of technologies (e.g., virtual reality technology, satellite imagery, and in-situ sensor data) are becoming available, providing opportunities to gain insights into challenging applied research areas such as environmental conservation. In this section, we describe new sources of data arising from subject elicitation using virtual reality and citizen science, as well as illustrating how Bayesian modelling can be applied to such data for the purposes of informing management decisions in Antarctica and the Australian Great Barrier Reef.    

\subsection{Elicitation using Virtual Reality}
Recent advancements in digital technologies have led to the large-scale collection of more advanced data such as sensor data, satellite imagery, and a host of varied resolution imagery and video including those taken using 360-degree cameras. It is possible, using these advances in technology, to enable location-specific data to remote researchers for analysis. The emergence of virtual reality (VR) technology, for example, acts as a way to connect the public and the scientific community, creating innovative pathways for environmental conservation research by immersing subjects in an otherwise inaccessible vivid virtual scene for elicitation purposes \cite{mazumdar2018citizen, queiroz2019immersive}. The opinions and knowledge extracted from this process is itself new data, which can be used for educational purposes \cite{fauville2020virtual} or incorporated into statistical models.

Increases in the volume of more complex types of data has led to the development of more effective and efficient analysis methodology. Recently, Bayesian models have seen use as a method to evaluate subject elicitation in the areas of coral reef conservation \cite{vercelloni2018using}, jaguar and koala habitat suitability assessments \cite{mengersen2017modelling, leigh2019using}, and the aesthetic value of sites in the Antarctic Peninsula.

\subsubsection*{Case Study: Quantifying Aesthetics of Tourist Landing Sites in the Antarctic Peninsula}

In the Antarctic Peninsula, the effects of climate change and the associated increase of ice-free areas are threatening the fragile terrestrial biodiversity \cite{lee2017climate}. As well as high ecological importance, these ecosystems also have a unique aesthetic value which has been formally recognised in Article~3 of the Protocol on Environmental Protection to the Antarctic Treaty \cite{parties1960protocol}. There is value in protecting beautiful landscapes, as tourism in Antarctica is based largely on the natural beauty of the environment. This case study quantifies aesthetic values in the Antarctic Peninsula by recording elicitation from subjects immersed in a virtual reality environment using a state-of-the-art web-based framework \verb|R2VR| \cite{vercelloni2021connecting}. 


Subject elicitation in this case study is drawn from $16$ photos, obtained via 360$^{\circ}$ photography at tourist landing sites in the Antarctic Peninsula. Consultation produced landscape characteristics of interest, e.g., the presence of certain animals and the weather. These characteristics and images were then used to construct an interview, to be held while the subject was immersed in the VR environment, with responses recorded on the Likert scale, from strongly disagree to strongly agree. From this elicitation process, responses to each question are recorded for each scene presented to the participant, as well as their opinion of the aesthetic value of the scene itself. Additionally, general participant characteristics such as gender identity and age are also recorded.

A Bayesian hierarchical model is used for modelling the response of whether or not a subject $i$ determines scene $j$ ($j=1,\ldots, o$) as aesthetically pleasing ($y_{ij}$) as a function of responses to statements such as ``there are animals in this image'' and ``this image is monotonous'' ($x_{ik}, k = 1,\dots, m$), subject characteristics such as age and gender ($x_{ih}, h = m,\dots, m+n$), and subject-reported confidence in their response to each interview statement ($s_{ij}, j=1,\ldots, m)$, where zero represents low confidence and one represents high confidence. The model is
\begin{align*}
y_{ij}|\alpha_j, \v \beta_{0 s_{ij}}, \v \beta_1,  &\stackrel{{\rm ind}}{\sim} \text{Bernoulli}\bigg({\rm logit}^{-1}\left(\alpha_{j} + \big(\v {\beta}_{0 s_{ij}}^\top, \v \beta_{1}^\top\big)^\top \v x_{i}\right)\bigg)\\
\v \alpha|\tau_{\alpha} & \ {\sim}\ \mathcal{N}(\v 0,\tau_{\alpha}^{-1}\m I_{o}), \\ 
\v \beta_{0l} \big|\v \mu, \v \tau_l & \stackrel{{\rm ind}}{\sim}\mathcal{N}\bigg(\v \mu, {\rm diag}\big(\v \tau_{l}^{-1}\big)\bigg), \quad l=0,1,\\
\v \beta_1  &\ {\sim}\ \mathcal{N}(\v 0, 10^{2} \m I_n), \\
\tau_{lk}, \tau_{\alpha} &\stackrel{{\rm iid}}{\sim} {\rm Gamma}(10^{-2},10^{-2}), \quad k = 1,\dots,m, \quad l=0,1 \\
\v \mu &\sim \mathcal{N}(\v 0, 10^2\, \m I_{m}).
\end{align*}

The development of conservation plans should, in accordance with the Protocol on Environmental Protection to the Antarctic Treaty, include recommendations based on aesthetic value. This case study is among the first to propose the incorporation of aesthetic value into conservation plans by leveraging subject-reported uncertainty. Understanding aesthetic attributes in Antarctica can be applied to other regions, especially through the implementation of similar surveys and models. The landscape of virtual reality data assets continues to expand as more researchers are made aware of the value added to methods of subject inquiry by including multi-modal features such as text, sounds, and haptic feedback. Modern Bayesian modelling approaches allow insights to be drawn from these such novel approaches to data collection.

\subsection{Citizen Science}
Citizen science (CS) represents one of the most popular emerging data sources in scientific research.
CS involves engaging members of the general population in one or more parts of the scientific process.   
Its applications can be found across almost all disciplines of science, especially in ecology and conservation where scientists are harnessing its power to help solve critical challenges such as climate change and the decline in species abundance. Examples of citizen scientists' contributions include reporting sightings of species, measuring environmental variables and identifying species on images.
Hundreds of citizen science projects can be found in popular online platforms including Zooniverse \cite{zooniverse}, eButterfly \cite{prudic2017ebutterfly}, eBird \cite{sullivan2009ebird}, and iNaturalist \cite{nugent2018inaturalist}. A fundamental issue often discussed surrounding citizen science is the quality of the data produced, which is generally error-prone and biased. For example, bias can arise in citizen science datasets due to (1) the unstructured nature of the data, (2) collecting data opportunistically, with more observations from frequently visited locations \cite{van2013occupancy} or at irregular frequencies across time \cite{dwyer2016using}, and (3) as a result of differing abilities of the participants to perform tasks such as detecting or identifying species \cite{strebel2014studying, santos2021MEE}. However, recent advances in statistics, machine learning, and data science are helping realize its full potential and increase trustworthiness \cite{freitag2016strategies, santos2020, santos2021MEE}.     

Frequently, citizen science data is elicited via image classification. For example, asking the participants whether images contain a target class or species. In this section, we illustrate two modelling approaches for these types of data. 

In the first approach, we consider a binary response variable $y_{ij}$ representing whether the category has been correctly identified by the participant ($i = 1,\ldots,m$) in the image ($j = 1,\ldots,n$).
The probability of obtaining a correct answer can be modeled using an item response model such as the three-parameter logistic model (3PL) \cite{baker2004item, santos2021MEE},
\begin{equation}
y_{ij}| Z_i, B_j, \eta_j, \alpha_j,  \sim {\rm Bernoulli}\big(\eta_j + \left(1 - \eta_j \right){\rm logit}^{-1}(\alpha_j(Z_i - B_j)\big),
\label {eq:eq11}
\end{equation}
where each $\eta_j \in (0,1)$ is a  pseudo-guessing parameter accounting for a participants' chance of answering correctly by guessing, $Z_i$ is the latent ability of the $i^{\textrm{th}}$ participant, $\alpha_j > 0$ is the slope parameter, and $B_j$ is the latent difficulty of the $j^{\textrm{th}}$ image.  
Sometimes, the correct answer for certain images is unknown. In this case, we estimate the latent labels for the images via the estimates of $Z_i$, by using the latter as weights in popular methods such as majority or consensus voting. 
Code to fit these models and exemplar datasets can be found in \cite{hakuna}.

The second approach is for the case where we are interested in the proportion of species in elicitation points in images. Here, we compute a statistic $\hat{y}_{ij} \in [0,1]$ giving the apparent proportion of species in a number of elicitation points in image $j$ classified by the participant $i$.
The true latent proportion $Y_j$ can be 
estimated based on each participant's overall performance measures~${\rm se}_i$ and~${\rm sp}_i$ (which denote the sensitivity and specificity scores of participant $i$, respectively). A Beta prior is placed on the true proportion, yielding the model
\begin{equation} 
\hat{y}_{ij} = Y_{j}\, {\rm se}_{i} + (1 - Y_{j})(1 - {\rm sp}_{i}),
\label{eq:yhat}
\end{equation} 
\begin{equation*} 
Y_j \sim \textrm{Beta}(\alpha_j,\beta_j ),
\end{equation*} 
where $\alpha_j$ and $\beta_j$ are the shape and the scale parameters in the beta distribution, respectively. 
The above model can be parametrized via a specified prior mean $\mu_j$ for each $Y_j$, and a common precision parameter~$\phi$, via ${\alpha_j = \mu_j \phi}$ and ${\beta_j = -\mu_j \phi + \phi}$, which in turn implies that
${\V [Y_j] = \frac{\mu_{j}(1-\mu_{j})}{(1+\phi)}}$. Covariates can also be incorporated by defining a beta regression with $\textrm{logit}(\mu_j) = \v \xi^{\top} \v x_j + U_j + \varepsilon_j$, where $\varepsilon_j$ are error terms, and  $U_j$ are spatially-dependent random effects.
Both approaches account for spatial variation (captured in  $B_j$ or $U_j$ for the first and second approach, respectively) using different spatial structures (e.g., conditional autoregressive (CAR) priors, covariance matrices, or Gaussian random fields).
See more details in \cite{santos2020, peterson2020monitoring, santos2021MEE}.

\begin{figure}
\centering
\begin{subfigure}{.5\textwidth}
  \centering
  \includegraphics[width=1.0\linewidth]{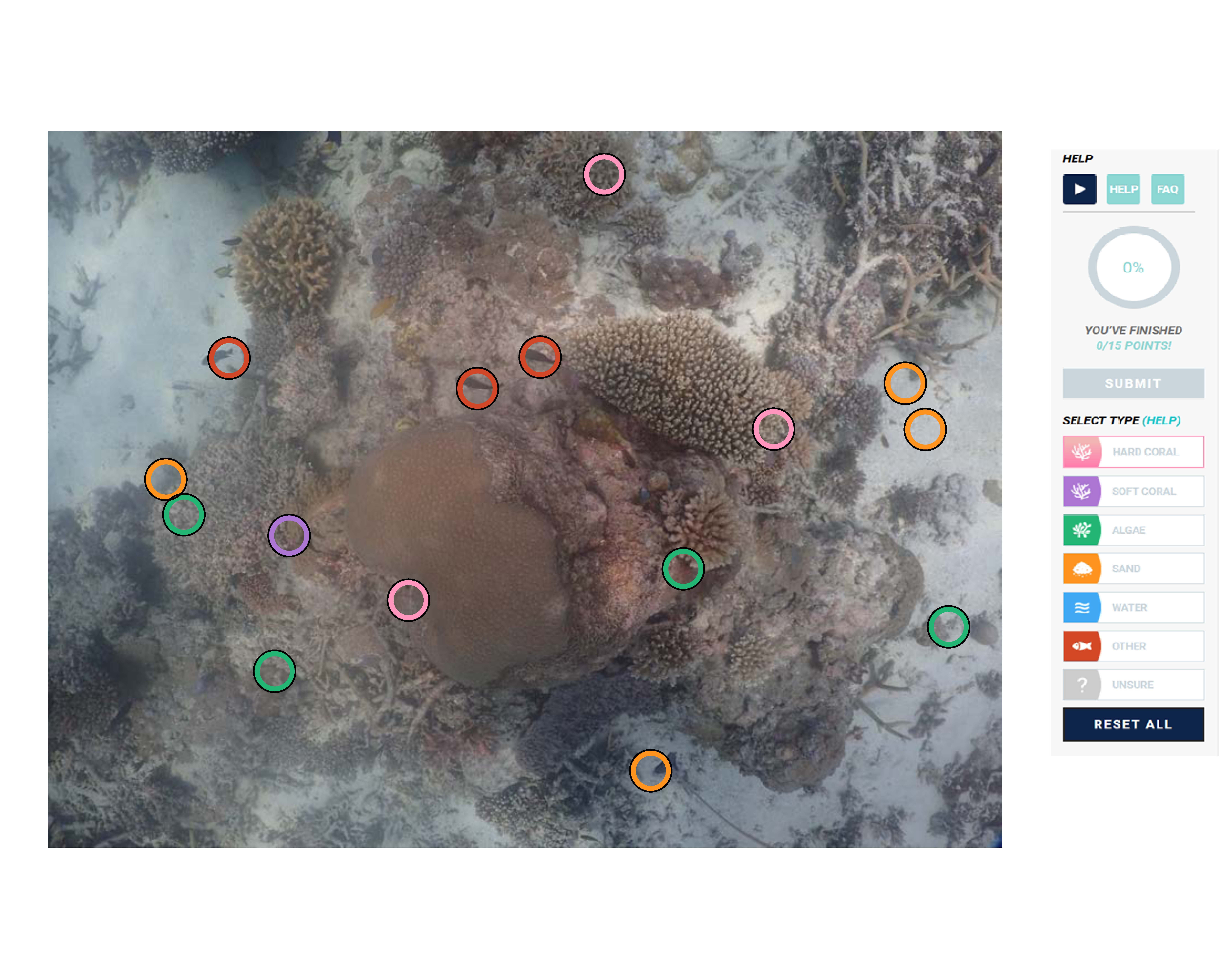}
   \label{fig:sub9}
\end{subfigure}%
\begin{subfigure}{.5\textwidth}
  \centering
  \includegraphics[width=1.0\linewidth]{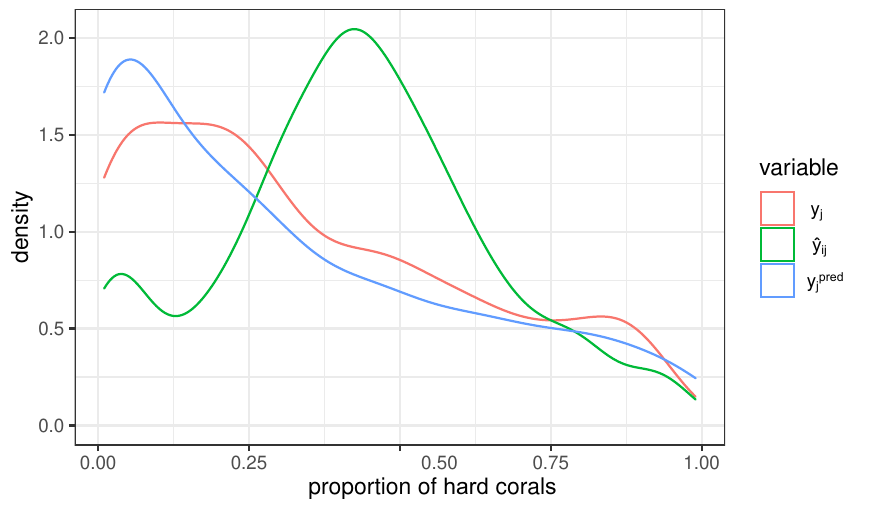}
   \label{fig:sub10}
\end{subfigure}       
\caption{(a) Elicited points with benthic categories in an underwater image from Great Barrier Reef, Australia.
(b) True latent proportion (in red) and the apparent proportion of hard corals (in green). The predicted proportion is represented in blue.}
\label{fig:reef}
\end{figure}

The following case study illustrates the estimation of the latent proportion of hard corals across the Great Barrier Reef in Australia, obtained from underwater images classified by citizen scientists. 
Figure \ref{fig:reef} shows 15 spatially balanced random points in one of the images used in the study. The apparent proportion of hard coral in the image was obtained using the number of points selected by participants containing this category out of 15. Using Equation \eqref{eq:yhat}, the (biased) estimates obtained from the citizen scientists can be corrected producing a similar density to the latent unobserved proportions.

The integration of citizen science data with current monitoring efforts from Australian federal agencies and non-governmental organisations is a breakthrough to increase the amount of information about changes along the Great Barrier Reef, learn about climate change impacts and adapt management actions consequently. This model introduced here is the root of a digital platform that estimates the health of the Great Barrier Reef using all available information. Australian Cancer Atlas (\url{atlas.cancer.org.au})  is a citizen science program that involves non-experts to collect information online and models it appropriately via a Bayesian approach. This study contributes to increasing the trust of citizen science and produce reliable data for environmental conservation while engaging and arising awareness about coral reefs.      

\section{Federated Analyses and Inference Methods}

\subsection{Overview}
In many areas of study (health, business, and environmental science, for example), the collective data set of interest one wishes to use in modelling is often under the control of different data custodians, i.e., parties responsible for ensuring data is only used or released in instances deemed appropriate to governance requirements.  Such requirements often stipulate that the data itself, and information pertaining to it can only be shared in a manner deemed sufficiently private. 

 Federated learning is the process of fitting a model in the setting where data resides with multiple custodians. Approaches typically place privacy as having the utmost importance, but computational efficiency is important from a practical perspective. There are two broad data settings that occur, each requiring their own style of algorithm. The first is the horizontal setting. Here, multiple data custodians have the same set of variables for different entities. In contrast, the vertical federated learning setting has data custodians who possess different variables for the same entities. An example of a horizontal setting is where different countries possess the data for those primarily residing within. In contrast, an example of vertical federated learning would be where two companies possess their respective sales data for the same collection of customers.
The term ``federated learning'' originated in the deep learning literature with the introduction of the \textit{FedAvg }algorithm \cite{mcmahan2017communication}. \textit{FedAvg} involves updating parameter values of a global model to be the weighted average of parameter values obtained by updating the same model locally (possibly many times) at each iteration. This work led to many related optimization algorithms, e.g., \textit{FedProx}\cite{li2020federated}, and \textit{FedNova }\cite{ wang2020tackling} which account for heterogeneous (non i.i.d.) data sources, and the Bayesian nonparametric approach for learning neural networks of \cite{yurochkin2019bayesian}, where local model parameters are matched to a global model via the posterior distribution of a Beta-Bernoulli process \cite{thibaux2007hierarchical}. To date, practical federated analyses appear restricted to the frequentist setting. Examples include the prediction of breast cancer using distributed logistic regression \cite{deist2020distributed} and modelling of the survival of oral cavity cancer through a distributed proportional Cox hazards model \cite{geleijnse2020prognostic}. Both these approaches conduct parameter estimation via a Newton-Raphson algorithm \cite{cellamare2022federated} and result in equivalent maximum likelihood estimates to those obtained in a standard, non-federated setting. Algorithms for the maximum likelihood estimation of log-linear and logistic regression models in vertical federated learning settings \cite{fienberg2006secure, slavkovic2007secure, shi2016secure, li2016vertical, kamphorst2022accurate} use ideas such as secure multiparty computation \cite{cramer2015secure}, and formulating the parameter estimation task as a dual problem \cite{minka2003comparison}. Several overarching software infrastructures such as \verb|VANTAGE6| \cite{moncada2020vantage6} ensure the correct and secure use of the data of each custodian within the specified algorithm, given acceptable (model- and application-specific) rules for information exchange.

Despite the potentially enabling capabilities of federated methods, to our knowledge, Bayesian federated learning methods have yet to impact real-world applications. In the Bayesian inference setting, the ``learning'' task becomes one of performing posterior inference, e.g., via Markov chain Monte Carlo (MCMC) or variational inference (VI) techniques. Note that Bayesian federated learning approaches may involve multiple communication rounds, though this is only sometimes the case. For example, many distributed MCMC approaches (e.g., \cite{wang2014parallelizing, neiswanger2014asymptotically, scott2016bayes}), combine individually-fit model posteriors, requiring only a single communication step from each local node. A recent intermediate approach \cite{jordan2019communication} is to construct a surrogate likelihood of the complete data set via an arbitrarily-specified number of communication steps. After constructing the surrogate likelihood, an MCMC algorithm is run on a single device. As the number of communication steps increases, the approximation error introduced by the surrogate likelihood decreases.
\begin{figure}[!ht]
\centering
\begin{subfigure}{.5\textwidth}
  \centering
  \includegraphics[width=.9\linewidth]{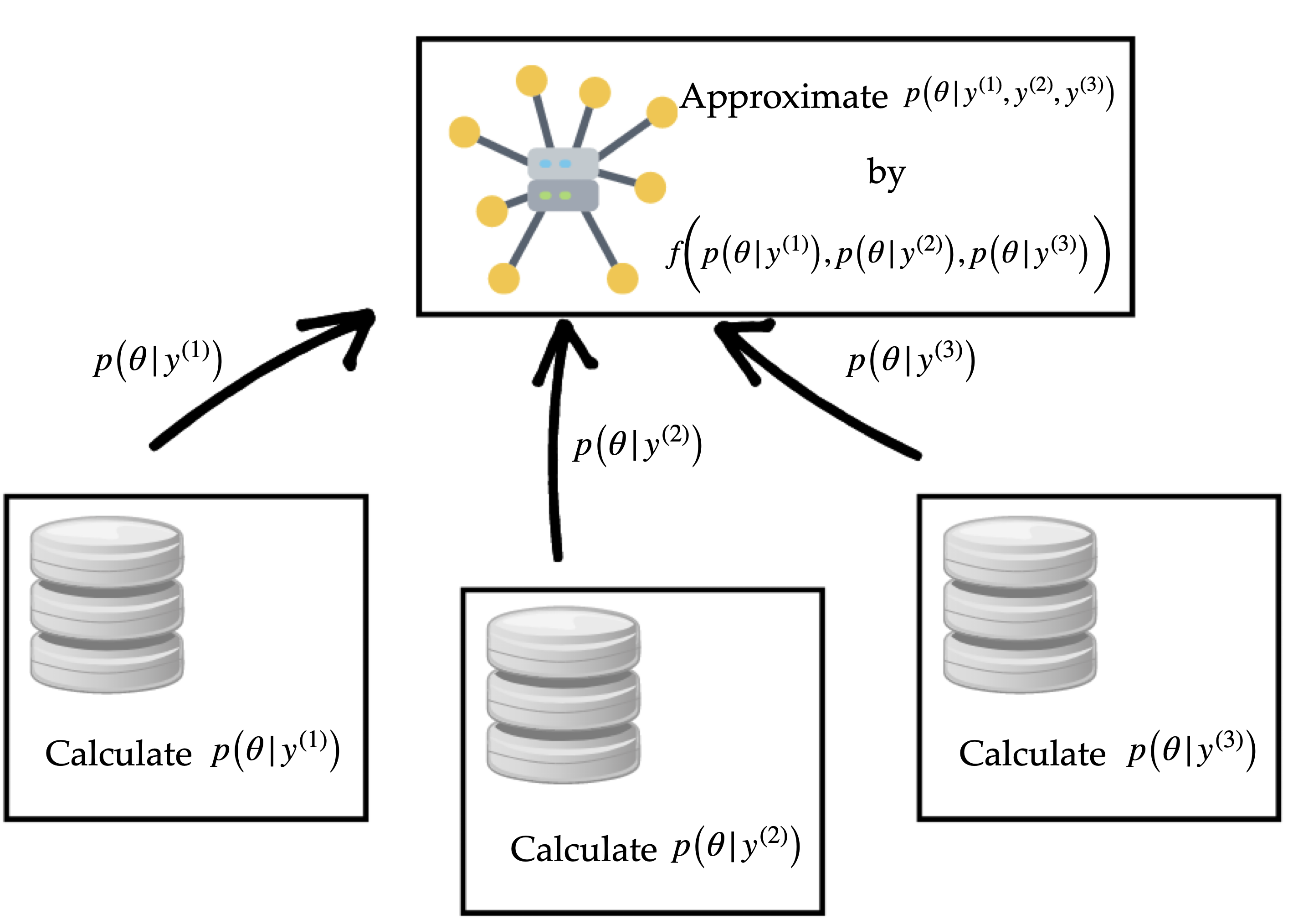}
   \label{fig:distributed_mcmc}
 \end{subfigure}%
\begin{subfigure}{.5\textwidth}
  \centering
  \includegraphics[width=.9\linewidth]{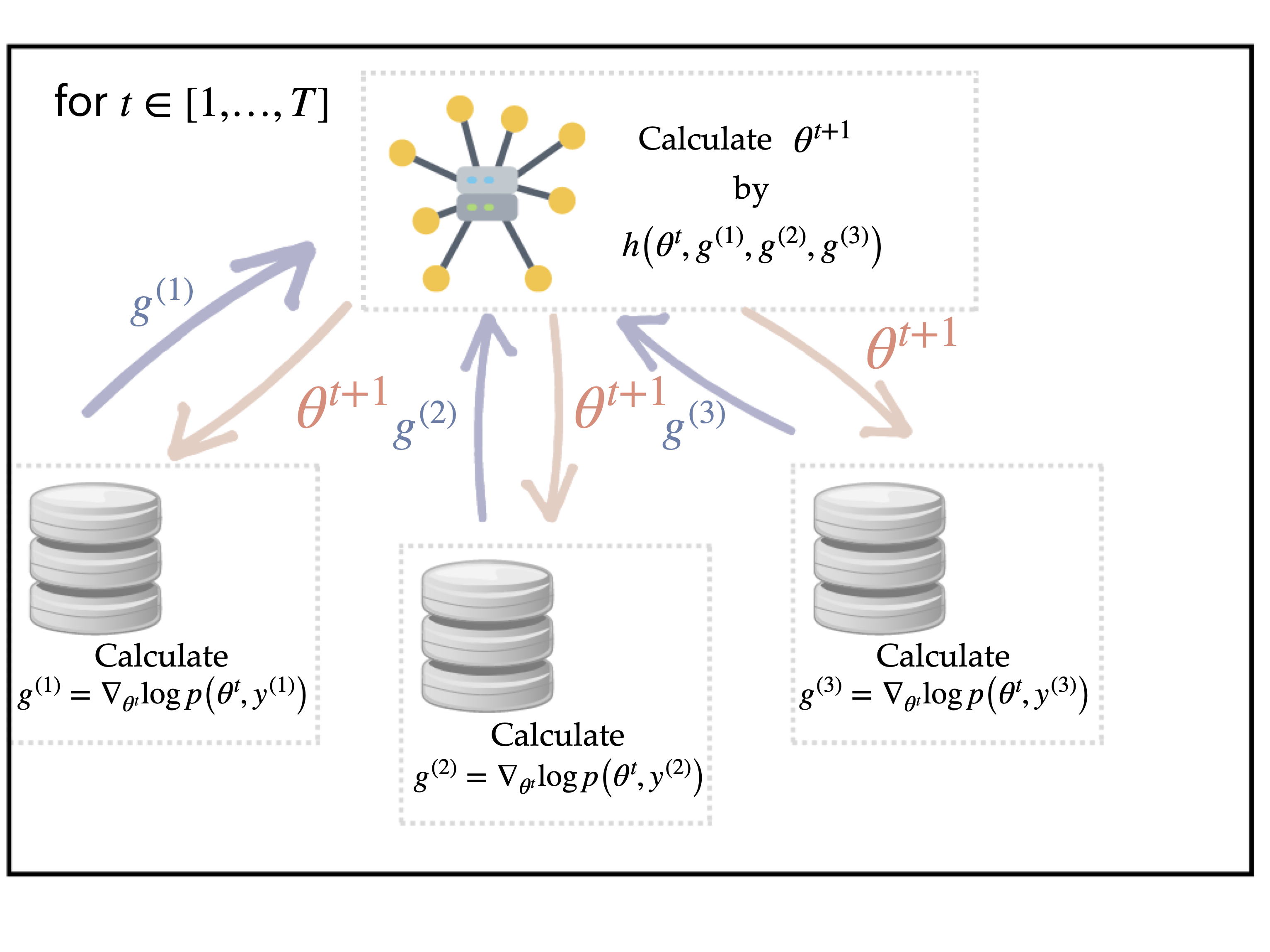}
   \label{fig:collab_learning}
\end{subfigure} 
\caption{Federated approaches lie on a continuum between post-hoc posterior amalgamation approaches as used in certain distributed MCMC approaches (left) and collaborative multi-round approaches (right).}
\label{fig:comparison_estimation}
\end{figure}

In certain cases, carrying out federated Bayesian inference is (at least in principle) relatively straightforward. For example, a naive MCMC algorithm would be trivial to construct for a simple model class, such as any generalized linear model (which assumes the data are independent), provided that one is not concerned with the number of communication steps. To see this, note that the (log-)posterior density function decomposes as
\begin{equation}\label{eq:fed_GLM}
    \log p(\v \theta| \v y) =  \log p(\v \theta) + \sum_{k=1}^n \log p(y_k |\v \theta)
 + {\rm
 const.}
\end{equation}

Hence, for the horizontal setting, all that is required is the nodes sharing the sum of their respective log-likelihood terms with the server. 
However, this approach would require a minimum of two communication steps per iteration of the Markov chain. Recent MCMC methods, similar in style to the \verb|FedAvg| algorithm (which use Langevin dynamics to update the Markov chain), require only a single communication step per iteration \cite{plassier2021dg, el2021federated}. Such approaches exploit gradient information which decomposes as a sum similarly to \eqref{eq:fed_GLM}, though eschew the usual Metropolis--Hastings correction and are hence asymptotically inexact. In some instances, a formally-justified notion of privacy may be required, as opposed to simply an intuitive one given by aggregation of terms. Differential Privacy (DP) (e.g.,\cite{de2020overview}) provides such guarantees, and there are variants of MCMC that ensure this, such as DP-MCMC \cite{heikkila2019differentially}, which accomplishes privacy guarantees at the cost of a slight perturbation of stationary distribution of the chain. It is worth noting that all of the above examples mentioned are specific to the horizontal setting, with the vertical setting proving especially challenging as one does not have a beneficial decomposition like that of \eqref{eq:fed_GLM}.

As the above alludes to, the development and use of Bayesian federated learning algorithms are complex for several reasons. A method is only suitable for a prescribed application if it satisfies a combination of requirements, such as being able to work with the desired model, computational and communication costs, privacy, and accuracy. For each application, the choice of model and federated method will depend on where the priorities lie, e.g., accuracy, efficiency, or privacy. In some cases, there may be no feasible algorithm (an example is given in the upcoming case study). Thus, inference approaches that improve upon some (or even all) of these aspects are important and warrant future research. 

The ultimate goal of federated Bayesian analysis is to circumvent the need for data merging  \cite{bohensky2010data, harron2017challenges} in scenarios where merging is considered infeasible. However, for Bayesian federated learning to reach this point, these approaches must offer custodians and interested parties an accurate inference for complex models while maintaining a level of privacy acceptable to those data custodians. Thus, the methodological development that enables federated inference for more advanced Bayesian models efficiently and/or with additional privacy guarantees is likely to emerge as a critical area of interest in the coming years.

\subsection{Case Study: Federated Learning with Spatially-Dependent Latent Variables}

The greatest hindrance to employing federated learning in real-world applications is the lack of possible model types that current algorithms address. Commonly, applied statistical modelling involves incorporating hierarchical structures, and latent variables \cite{gelman2006data}. To our knowledge, there are no federated Bayesian analysis algorithms {\em at all} for such models. To briefly illustrate the unique challenges and the need for developments that account for the nuances of different models, the case study considers spatially-dependent latent variables based on neighbourhood structures. For simplicity, the focus is on the Intrinsic Conditional AutoRegressive (ICAR) prior \cite{besag1974spatial}, although variations such as the Besag--York--Mollie  \cite{besag1991bayesian} and Leroux \cite{leroux2000estimation} models are similar in what follows (the latter is used for example, in the Australian Cancer Atlas described in Section \ref{sec:purposefulprods}). The ICAR prior posits a vector of spatially-dependent latent variables, denoted here as $\v Z$. Each element of $\v Z$ corresponds to a latent area-level effect of a ``site'', which is influenced by neighbouring sites. Writing $i\sim j$ to denote that sites $i$ and $j$ are considered neighbours, and assuming the graph arising from the neighbourhood structure is fully connected, the ICAR prior with precision hyperparameter $\tau$ has log-density 
\[\log p(\v z; \tau) = \frac{n}{2}\log \tau  -\frac{\tau}{2} \sum_{i \sim j} (z_i - z_j)^2 + {\rm const.}\] 
The above may be problematic if the data custodians insist that the latent variables corresponding to their areas must be kept private to themselves. To see why, consider the case that there are two (2) data custodians, with the sets ${\cal C}_1$ and ${\cal C}_2$ containing the indices of data possessed by the first and second custodian, respectively. Then,
\begin{equation}\label{eq:fed_crossterm} 
\sum_{i \sim j} (z_i - z_j)^2 = \sum_{i,j \in {\cal C}_1:i \sim j} (\textcolor{blue}{z_i} - \textcolor{blue}{z_j})^2  + \sum_{i,j \in {\cal C}_2:i \sim j} (\textcolor{red}{z_i} - \textcolor{red}{z_j})^2  + \sum_{i \in {\cal C}_1, j \in {\cal C}_2 : i\sim j} (\textcolor{blue}{z_i} - \textcolor{red}{z_j})^2,
\end{equation}
where terms in \textcolor{blue}{blue} are those relevant to the sites under the first custodian, and those in \textcolor{red}{red} to the second. When computing the log-posterior density (as required, for example, in Markov chain Monte Carlo algorithms), the first two terms on the right-hand side above can be aggregated and sent to the central server. However, the final term can not as each individual summand requires the individual latent variables to be processed. This is because the latter term considers interactions across custodian boundaries. 

Consequently, solutions such as (i) employing judicious reparameterization of the latent variables (possibly compatible with the one that is often also required to enforce identifiability), (ii) changing the model to add additional auxiliary variables, or (iii) otherwise approximating the troublesome term, are required. An additional challenge is that even if inferences on individual latent variables are only available to their respective custodians, they may nevertheless ``leak'' information across custodian boundaries to neighbouring sites due to the underlying dependency structure.

While the above certainly highlights particular challenges, the first two terms of \eqref{eq:fed_crossterm} split nicely across custodians and hint that latent variables need not always be problematic. For more straightforward cases such as the latter, specialized accurate and efficient inference approaches that allow individual custodians to avoid ever sharing their latent variables (either directly or indirectly) or data are the subject of forthcoming work by the authors of this section, who have a longer-term goal of tackling more challenging cases such as ICAR and its relatives in different settings.

\section{Bayesian Inference for Implicit Models}

\subsection{Overview}

The appetite for developing more realistic data-driven models of complex systems is continually rising.  Development of such complex models can improve our understanding of the underlying mechanisms driving real phenomena and produce more accurate predictions.  However, the calibration of such models remains challenging, as the associated likelihood function with complex models is often too computationally cumbersome to permit timely statistical inferences.  Despite this, it is often the case that simulating the model is orders of magnitude faster than evaluating the model's likelihood function.  Such models with intractable likelihoods that nevertheless remain simulable are often referred to as implicit models.  Such models are now prevalent across many areas of science (see e.g., various application chapters in \cite{Sisson2018}).

Currently, the most popular statistical approach amongst practitioners for performing Bayesian inference for implicit models is approximate Bayesian computation (ABC), popularised by \cite{Beaumont2002}.    A related method called generalised likelihood uncertainty estimation \cite{beven1992future,beven2014glue} predates ABC, and \cite{nott2012generalized} explore its connections to ABC.   The ABC approach approximates the true posterior as
\begin{align}
    p_\epsilon(\boldsymbol{\theta}|\textbf{y}) & \propto \pi(\boldsymbol{\theta})\int p(\textbf{x}|\boldsymbol{\theta})\mathbb{I}(||\textbf{x}-\textbf{y}|| \leq \epsilon)d\textbf{x}. \label{eq:ABC_posterior}
\end{align}
Here, $\textbf{x}$ denotes simulated data that has the same structure as $\textbf{y}$, $||\cdot||$ is some norm (i.e., $||\textbf{x} - \textbf{y}||$ measures the closeness of the simulated data to the observed data), and $\epsilon$ stipulates what is considered ``close''.  Intuitively, values of $\boldsymbol{\theta}$  more likely to produce simulated data $\textbf{x}$ close enough to $\textbf{y}$ have increased (approximate) posterior density.  Rather than compare $\textbf{y}$ and $\textbf{x}$ directly, it can be more efficient to compare $\textbf{y}$ and $\textbf{x}$ in a lower dimensional space via a summarisation function that aims to retain as much information from the full data set as possible.    For the posterior in \eqref{eq:ABC_posterior} to equate to the exact posterior, we require that the observed and simulated datasets are matched perfectly (i.e.,\ as $\epsilon \rightarrow 0$) in terms of some sufficient summarisation. However, in the majority of practical applications, a low-dimensional sufficient statistic does not exist and it is computationally infeasible to take $\epsilon \rightarrow 0$, so we must accept some level of approximation.

Given the wide applicability of the approach, i.e., that only the ability to simulate the model is required to conduct inference, there has been an explosion of research in the past 10--15 years advancing ABC and related methods that lie within the more general class of so-called likelihood-free inference methods.  A substantial portion of methodologically-focused ABC research considers aspects including the effective choice of $||\cdot||$ (e.g., \cite{prangle2018summary,drovandi2022comparison}), efficient sampling algorithms  to explore the approximate posterior in \eqref{eq:ABC_posterior} (e.g.,\ \cite{Sisson2018a}) and ABC's theoretical properties (e.g.\ \cite{frazier+mrr18}).  Many of the developments of ABC and some related methods (e.g.\ Bayesian synthetic likelihood \cite{Price2018,frazier2020BSLasymp}) prior to 2018 are discussed in \cite{Sisson2018}, the first-ever monograph on ABC.

The following case study considers a popular class of sampling algorithms for ABC based on sequential Monte Carlo (SMC).  SMC-based ABC algorithms improve efficiency compared to sampling naively from the prior by gradually reducing the ABC tolerance $\epsilon$ where the output produced at iteration $t$ is used to improve the proposal distribution of $\boldsymbol{\theta}$ at iteration $t+1$.  The output of the algorithm is $N$ samples, or ``particles'', from the ABC posterior in \eqref{eq:ABC_posterior} with a final $\epsilon$ that is either pre-specified or determined adaptively by the algorithm.  Each particle has attached to it a ``distance'', which is the value of $||\textbf{x}-\textbf{y}||$ for $\textbf{x}$ simulated from the model based on the particle's parameter value.     Here, we use the adaptive SMC-ABC algorithm in \cite{carr2021estimating}, itself a minor modification of the replenishment algorithm of \cite{Drovandi2011b}.  The algorithm is summarised below.
\begin{enumerate}
\item Draw $N$ samples from the prior, and for each sample, simulate the model and compute the corresponding distance. Initialise $\epsilon$ as the largest distance amongst the set of particles.
\item Set the next $\epsilon$ as the $\alpha$-quantile of the set of distances.  Retain the $N_\alpha$ particles with distance less than or equal to $\epsilon$.
\item Resample the retained particle set $N-N_\alpha$ times so that there are $N$ particles.
\item Run MCMC on each of the resampled $N-N_\alpha$ particles with stationary distribution \eqref{eq:ABC_posterior} with the current $\epsilon$.  This step helps to remove duplicate particles created from the previous resampling step.  The number of MCMC iterations can be adaptively set based on the MCMC acceptance rate.  
\item Repeat steps (ii)-(iv) until a desired $\epsilon$ is reached or the MCMC acceptance rate in (iv) is too small (i.e., the number of MCMC steps becomes too large for the computational budget).
\end{enumerate}
A key computational inefficiency of ABC and closely-related methods such as BSL is that many of the model simulations yield MCMC proposals that  are rejected. To obtain a suitable quality of approximation, it is not uncommon to require continuing the algorithm past the point where $\epsilon$ is small enough to have average acceptance probabilities of $10^{-2}$ or less.  To overcome this issue, there has been significant attention devoted to machine learning based approaches to likelihood-free inference, especially in the past five years.  These methods use model simulations (from different parameters) as training data for building a conditional density estimator of the likelihood (e.g., \cite{papamakarios2019sequential}), likelihood ratio (e.g., \cite{thomas2022likelihood}) or posterior density (e.g., \cite{lueckmann2017flexible}). Following this estimation, standard methods from the Bayesian inference toolkit can be used. Many machine learning approaches to likelihood-free inference can be implemented sequentially, so that samples from the approximate posterior in the previous (or all previous) iterations can comprise an increasingly informed training set that yields a more accurate conditional density estimator in regions of non-negligible posterior probability.  For the case study below, we compare the SMC-ABC approach with the sequential neural likelihood (SNL) method of \cite{papamakarios2019sequential}, which is outlined below.
\begin{enumerate}
    \item Set the initial proposal distribution of parameter values as the prior, i.e.,\ $q(\boldsymbol{\theta}) = p(\boldsymbol{\theta})$.
    \item Generate a training data set by  drawing $M$ parameter/simulated data pairs according to $q(\boldsymbol{\theta})p(\mathbf{x}|\boldsymbol{\theta})$.  Fit a conditional normalising flow (a flexible type of regression-density estimator) to the training data to estimate the conditional density of $\mathbf{X}|\boldsymbol{\theta}$.
    \item Run MCMC to obtain approximate posterior samples, using the learned conditional density of $\mathbf{X}|\boldsymbol{\theta}$ evaluated at the observed data $\mathbf{y}$ as the approximation to the likelihood.  Samples from this approximate posterior can also be used to update the proposal distribution $q(\boldsymbol{\theta})$.
    \item Repeat steps (ii) and (iii) for a desired number of rounds.
\end{enumerate}

\subsection{Case Study: Calibrating Agent-based Models of Tumour Growth}

In this case study, we apply likelihood-free methods SMC-ABC and SNL for calibrating a complex agent-based model (ABM) of tumour growth.  We briefly compare the methods in terms of computationally efficiency and their ability to fit simulated and real tumour growth data. 

ABMs have been used in cancer modelling for some time now as they provide a spatial representation of the inherent cellular heterogeneity and stochasticity of tumours \cite{wang2015simulating,metzcar2019review,macnamara2021biomechanical}. Largely, these models account for the individual cell based behaviours of proliferation, movement and death and aim to predict the impact of stochasticity on spatial tumour growth over time. Previous works have considered this in the context of angiogenesis \cite{cess2022multiscale}, immune involvement \cite{norton2019multiscale,cess2020multi} and also treatment \cite{jenner2022agent}. In some cases, data has been used to calibrate or validate aspects of the models \cite{klowss2022stochastic,gallaher2020cells}, although due to the computational cost and their intractable likelihood it is not always easy to infer parameters in an ABM using data. 

For this case study, we use a previously published ABM called a Voronoi Cell-Based model (VCBM) \cite{jenner2022examining,jenner2020enhancing}. In this model, cancer cells and healthy tissue cells are considered agents, whose centre is modelled by a point on a 2D lattice, and whose boundary is defined by a Voronoi tessellation. To mimic tumour growth and spatial tissue deformation, the model captures cell movement using force-balance equations derived from Hooke's law. In this way, cell movement is captured off-lattice and is a function of the local cell-neighbourhood pressure, determined using a Delaunay Triangulation. 

Tumour growth is captured by introducing a probability of an individual cancer cell proliferating $P$, which is a function of a cell's distance to the boundary of the tumour: $P = p_0\left(1-\frac{d}{d_{\mathrm{max}}}\right)$,
where $p_0$ is the probability of proliferation, $d$ is the cell's Euclidean distance to the tumour boundary (measured from the cell centre to the nearest healthy cell centre) and $d_{\mathrm{max}}$ is the maximum radial distance a cell can be from the boundary and still proliferate. In this way, the model evolves stochastically over time with cells either proliferating or moving in a given timestep. The model also uses $g_{\mathrm{age}}$ to define the time taken for a cell to be able to proliferate and uses $p_{\mathrm{psc}}$ as the probability of cancer cell invasion. Hence, the model parameter $\boldsymbol{\theta}$ to be estimated is $(p_0,p_{\mathrm{psc}},d_{\mathrm{max}},g_{\mathrm{age}})$.

\begin{figure}
\centering
\begin{subfigure}{0.5\textwidth}
    \begin{subfigure}{1\textwidth}
    \caption{SMC-ABC}
        \centering
        \includegraphics[width=.9\linewidth]{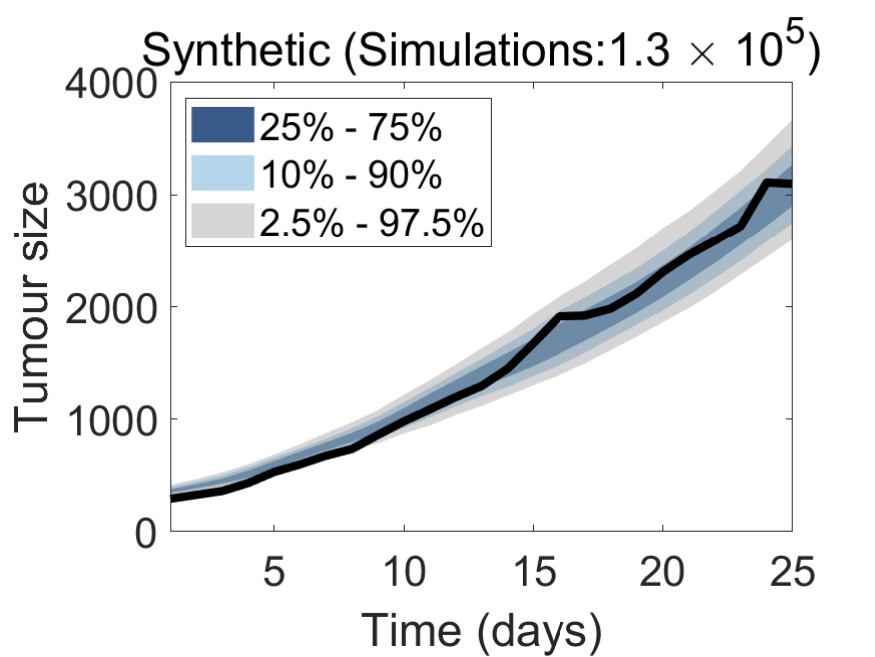}
        \label{fig:sub1}
    \end{subfigure}%
    \vfill
    \begin{subfigure}{1\textwidth}
        \centering
        \includegraphics[width=.9\linewidth]{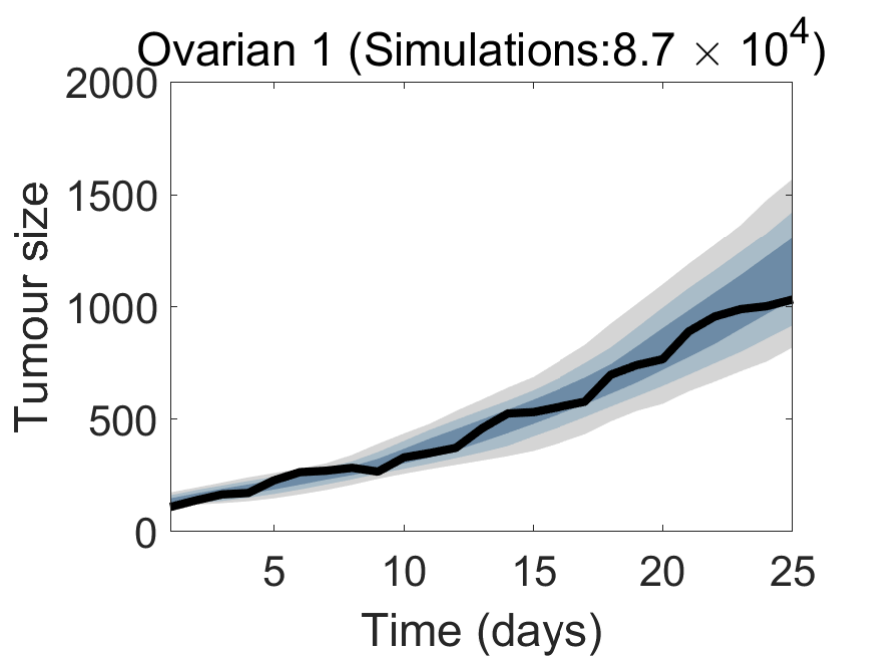}
        \label{fig:sub2}
    \end{subfigure}%
    \vfill
    \begin{subfigure}{1\textwidth}
        \centering
        \includegraphics[width=.9\linewidth]{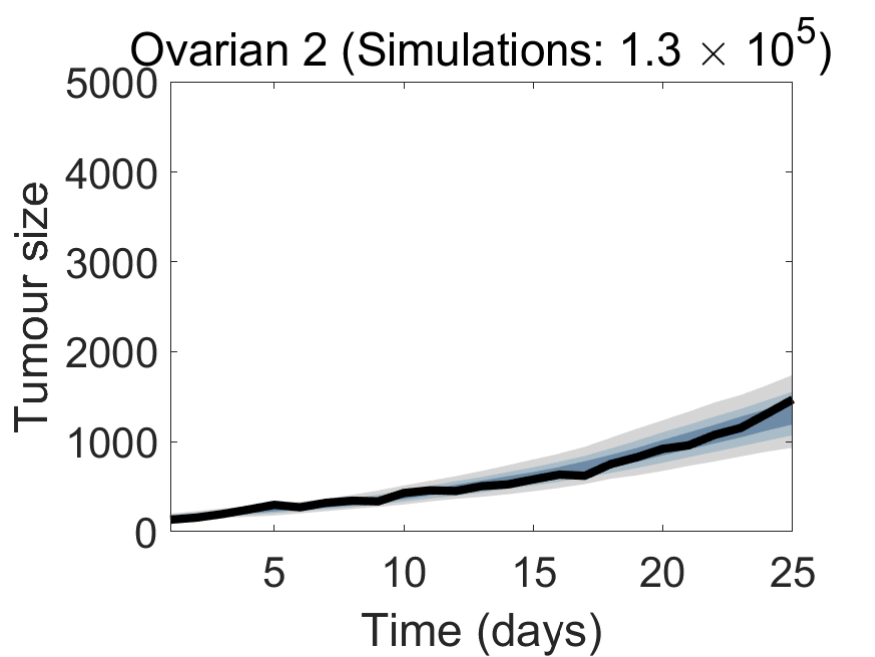}
        \label{fig:sub3}
    \end{subfigure}%
    \vfill
    \begin{subfigure}{1\textwidth}
        \centering
        \includegraphics[width=.9\linewidth]{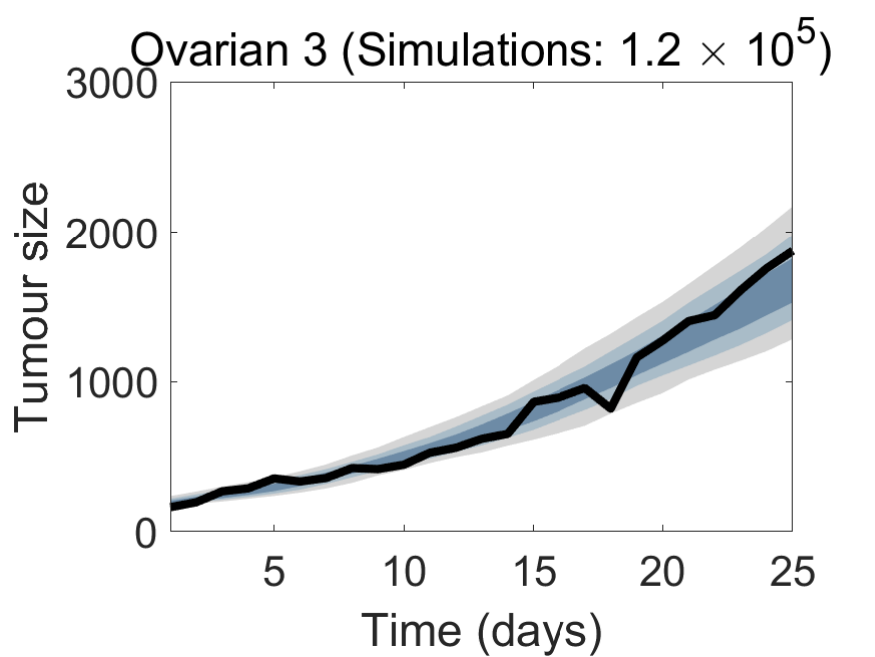}
        \label{fig:sub4}
    \end{subfigure}%
\end{subfigure}%
\hfill
\begin{subfigure}{0.5\textwidth}
\caption{SNL} 
    \begin{subfigure}{1\textwidth}
        \centering
        \includegraphics[width=.9\linewidth]{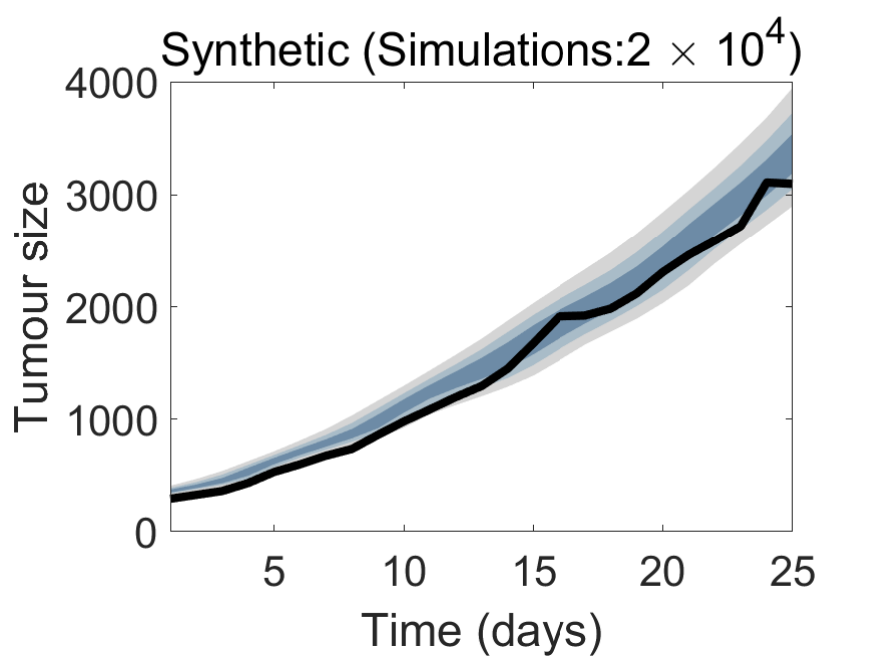}
        \label{fig:sub5}
    \end{subfigure}%
    \vfill
    \begin{subfigure}{1\textwidth}
        \centering
        \includegraphics[width=.9\linewidth]{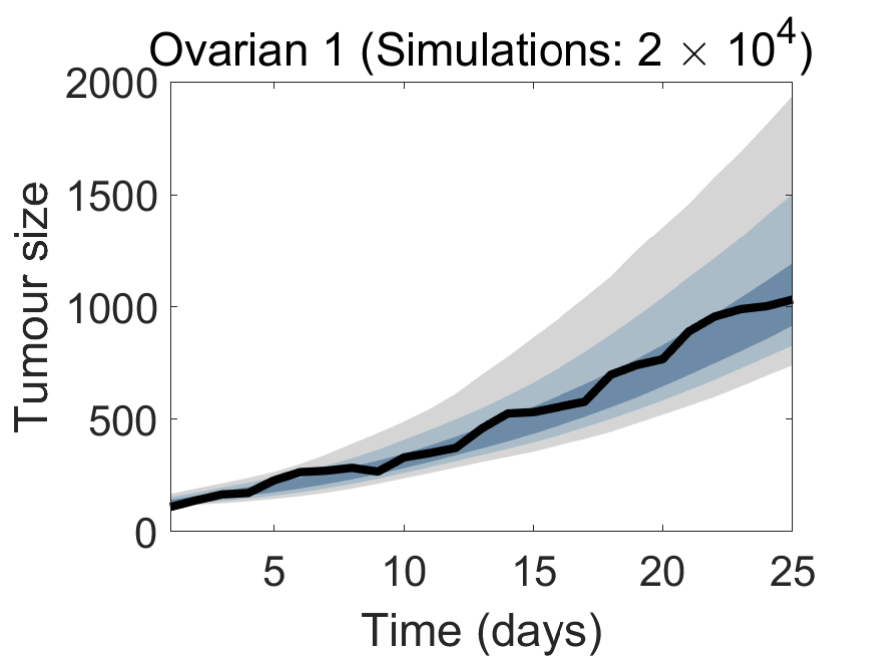}
        \label{fig:sub6}
    \end{subfigure}%
    \vfill
    \begin{subfigure}{1\textwidth}
        \centering
        \includegraphics[width=.9\linewidth]{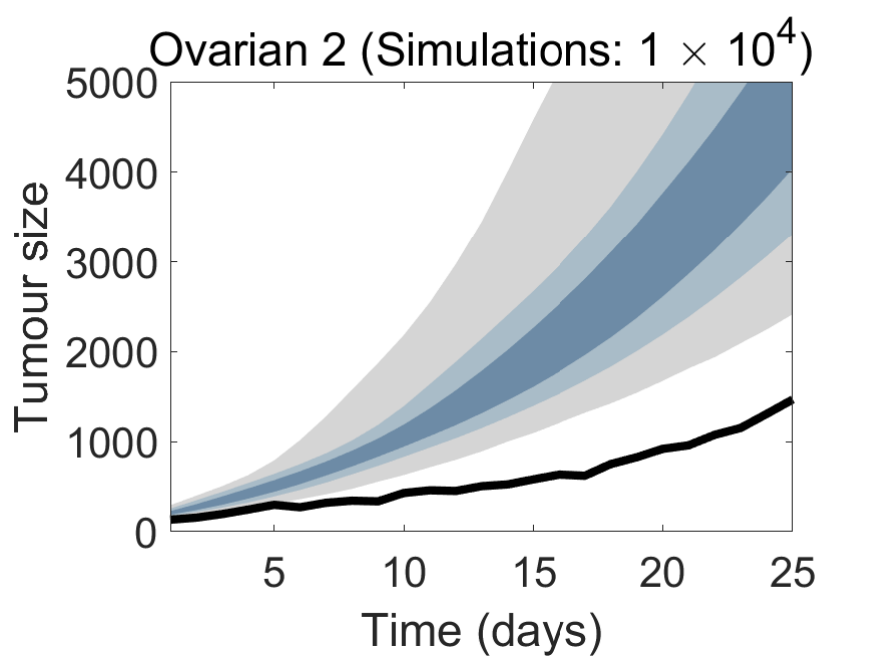}
        \label{fig:sub7}
    \end{subfigure}%
    \vfill
    \begin{subfigure}{1\textwidth}
        \centering
        \includegraphics[width=.9\linewidth]{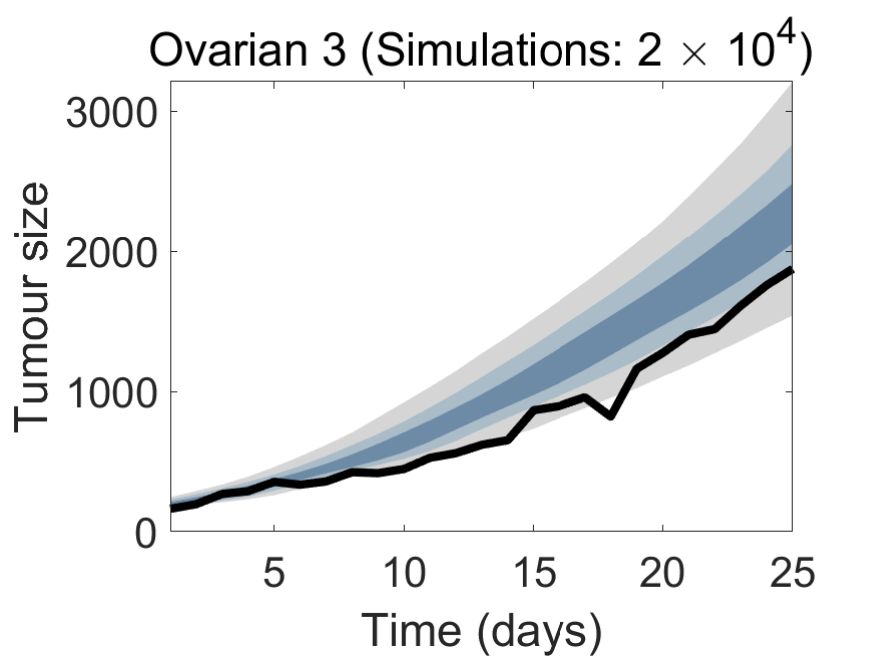}
        \label{fig:sub8}
    \end{subfigure}%
    
\end{subfigure}%
\caption{The posterior predictive distributions of (a) SMC-ABC and (b) SNL for the synthetic and ovarian cancer datasets. The black solid line is the tumour growth data. }
\label{fig:abm_results}
\end{figure}

To validate the VCBM, we use published in vivo tumour growth measurements for ovarian cancer \cite{kim2011active}. In these experiments, tumour volume was recorded by measuring the tumour width and length as perpendicular axis using calipers and then calculating the approximate tumour volume. We simulate the VCBM in 2D and calculate the corresponding tumour volume measurements equivalently.  We also consider one simulated dataset generated with parameter value $\boldsymbol{\theta} = (0.2,{10}^{-5},31,114)$.  The datasets are shown as solid black lines in Figure \ref{fig:abm_results}.  The prior distribution on $\boldsymbol{\theta}$ is given by $p_0 \sim \mathrm{Beta}(1,1)$, $p_{\mathrm{psc}} \sim \mathrm{Beta}(1,{10}^4)$, $d_{\mathrm{max}} \sim \mathrm{LogNormal}(\mathrm{log}(30),1)$ and $g_{\mathrm{age}} \sim \mathrm{LogNormal}(\mathrm{log}(160),1)$, with parameters assumed independent a priori \cite{wang2022calibration}.

We run SMC-ABC until around 100,000 model simulations have been generated for each dataset.  We use the SBI package \cite{tejero2020sbi} to implement SNL with five rounds of 10,000 model simulations for each dataset. To compare the performance of SMC-ABC and SNL, we compute the posterior predictive distribution for each dataset.  For SNL, we choose the round that visually produces the most accurate posterior predictive distribution.  We find that for SNL the performance can degrade with increasing rounds in three ovarian cancer datasets. 

The results are shown in Figure \ref{fig:abm_results}.  It can be seen that SMC-ABC produces posterior predictive distributions that tightly enclose the time series of tumour volumes for three real-world ovarian cancer datasets.   It is evident that SNL produces an accurate posterior predictive distribution for the synthetic dataset, with substantially fewer model simulations than that used for SMC-ABC.  This result is aligned with other synthetic examples in the literature (e.g., \cite{lueckmann2021benchmarking}). However, the SNL results for the real data are mixed, and for the three real datasets SMC-ABC produces more accurate posterior predictive distributions.  Further, we do not necessarily see an improvement in SNL when increasing the number of rounds (i.e., \ number of model simulations).  We suggest a reason for the potential poor performance of SNL is that the real data are more noisy than than what the simulator is able to produce, and this may lead to a poorly estimated likelihood generated by SNL when evaluated at the observed data. In contrast, SMC-ABC produces results that are more robust to this misspecification, albeit at a higher computational cost in terms of model simulations.  The potential poor performance of SNL under misspecification, and possible remedies to this problem, require further research.  In terms of computational cost, SMC-ABC takes $\sim$3 hours for each dataset.  For SNL, it takes $\sim$5 minutes to generate model simulations, $\sim$20 minutes to train the conditional normalising flow and $\sim$3 hours to generate approximate posterior samples (using the slice sampler as in \cite{papamakarios2019sequential}) for each round. The \textsf{C++} and \textsf{Python} codes used in this study are
available at \cite{ABCSNLgit}.

\section{Model Transfer}

Updating prior beliefs based on data is a core tenet of Bayesian inference. In the Bayesian context, model transfer extends Bayesian updating by incorporating information from a well-known source domain into a target domain. Consider the scenario where a target domain has insufficient data $y_{\mathcal{T}}$ to enable useful inference. Model transfer allows us to borrow information from a source domain with sufficient data $y_{\mathcal{S}}$ to improve inference. The transferability problem then is a question of when to transfer information, which information to transfer, and how to transfer this information. This problem appears across several domains, with some solutions exploiting the underlying properties of the source model, while others create informative priors with the source information. Below, we will discuss several different approaches to the model transfer problem. This broad topic is also known as transfer learning in the machine learning literature \cite{zhuang2020comprehensive}.

Naive updating, which uses all available source information, is a natural starting point to approach model transfer, though it can be detrimental. If the source and target distributions are dissimilar, negative transfer \cite{agarwal2021transfer} may occur reducing the inference or predictive power from our posterior. Power priors \cite{YE202229} correct for the  difference between source and target distributions by flattening the likelihood of the source distribution. This flattening is done by choosing a value~$\phi \in [0, 1]$ and raising the source likelihood to the value of $\phi$ which gives 
$$\pi(\boldsymbol{\theta}| \phi, y_{\mathcal{T}}, y_{\mathcal{S}}) \propto f_{\mathcal{T}}(y_{\mathcal{T}}|\boldsymbol{\theta})f_{\mathcal{S}}(y_{\mathcal{S}}|\boldsymbol{\theta})^{\phi} \pi(\boldsymbol{\theta}),$$
where $f_{\mathcal{S}}(y_{\mathcal{S}}|\boldsymbol{\theta})$ and $f_{\mathcal{T}}(y_{\mathcal{T}}|\boldsymbol{\theta})$ are the source and target likelihood functions, respectively. Naive updating would simply use the value $\phi = 1$. Finding an appropriate value for $\phi$ is challenging, intuitively we want to treat this as a latent variable and assign an appropriate prior. Unfortunately, even when both datasets are from the same distribution, the resulting posterior marginal of $\phi$ may exhibit only slightly less variance than the chosen prior. This phenomenon is analysed in \cite{pawel2022normalized} with illustrative examples. Other approaches attempt to determine an appropriate value of $\phi$ by optimisation. Different information criteria, from the standard deviance information criterion to more complex penalized likelihood-type criterion, have been used \cite{ibrahim2015power} including the marginal likelihood \cite{han2022study} and the pseudo-marginal likelihood \cite{bennett2021novel} which are evaluated using only the target data.

The transfer learning literature has a large number of methods for model transfer, evident by the recent review paper \cite{zhuang2020comprehensive}. Many of these methods are specific to neural networks, but some can still be applied to broader classes of statistical models. An example of such a method is described in \cite{app10020559} which uses an ensemble of convolutional neural networks with a majority voting selection step that is easily generalised for use beyond neural networks. Another method, TrAdaBoost.R2 \cite{gupta2022boosting,tang2020improving} adapts boosting \cite{solomatine2004adaboost} to the model transfer problem. This method iteratively reweights each data point in the source and target domain to improve the predictive performance of the target model. There are also several methods specific to generalised linear models. These use a variety of approaches to achieve model transfer for generalised linear models including; knockoff filters \cite{li2021transfer} to identify a subset of the source data to use, scaling the source likelihood function \cite{maity2021linear, reeve2021adaptive}, and regularization \cite{guo2017towards,hector2022turning} to adjust the weight of the source data. Finally, Transfer Gaussian Processes \cite{da2019fast,wei2022transfer,cao2010adaptive} attempt to use information from the source kernel to improve model performance on the target domain. This is achieved by pooling the source and target datasets and producing a new joint kernel 
$$\tilde{k}(\boldsymbol{x}, \boldsymbol{x}') =
\begin{cases}
\lambda k(\boldsymbol{x}, \boldsymbol{x}'), \qquad \mbox{if } \boldsymbol{x} \mbox{ and } \boldsymbol{x}' \mbox{ are in different domains} \\
k(\boldsymbol{x}, \boldsymbol{x}'), \qquad \ \mbox{ otherwise}
\end{cases}.$$
Above, $\lambda \in [0, 1]$, where $\lambda =0$ indicates no information transfer and $\lambda=1$ complete information transfer. For the interested reader, exemplar code is available via \cite{tgpgit}. 

Current state-of-the-art Bayesian model transfer generalises naive Bayesian updating but relies on fixed levels of transfer rather than incorporating uncertainty. It is still not clear how one should learn an optimal $\phi$ value in this paradigm but we expect future research will address this and use uncertainty more effectively. Moreover, given the interest in model-specific transfer learning, we believe that a Bayesian approach will be useful to develop general methods that are model agnostic.

\section{Purposeful products}\label{sec:purposefulprods}
A key advantage of Bayesian methods is their ability to assist in decision making, and here three different case studies showcase innovative tools using Bayesian approaches. 
\subsection{CoRiCAL: COVID-19 Vaccine risk-benefit calculator}
During the first year of the COVID-19 pandemic in 2020, border closures, lock-downs and other favourable conditions meant that Australia was spared from the high per capita case numbers and COVID-19 related deaths that were experienced in many other countries. When vaccines became available in February, 2021 \cite{gov}, the low number of COVID-19 related fatalities in Australia was coupled with uncertainty around highly publicised rare adverse-events for the vaccines; Thrombosis and Thrombocytopenia Syndrome (TTS) from AstraZeneca \cite{greinacher2021thrombotic} and myocarditis from Pfizer \cite{marshall2021symptomatic}. This led to high levels of vaccination hesitancy in the general public \cite{leask2021communicating}.  Although emerging scientific evidence was increasingly available on the risks of the vaccines and their effectiveness against both becoming infected and becoming severely ill once infected \cite{sheikh2021sars} \cite{zheng2022real}, compiling and assessing this information from Scientific journals and Government reports is impossible for the majority of the population. Collating this evidence into an easily understood format that could be used by people to make an informed decision on COVID-19 vaccination in the Australian context became crucial. 

\begin{figure}
\includegraphics[width=1\textwidth]{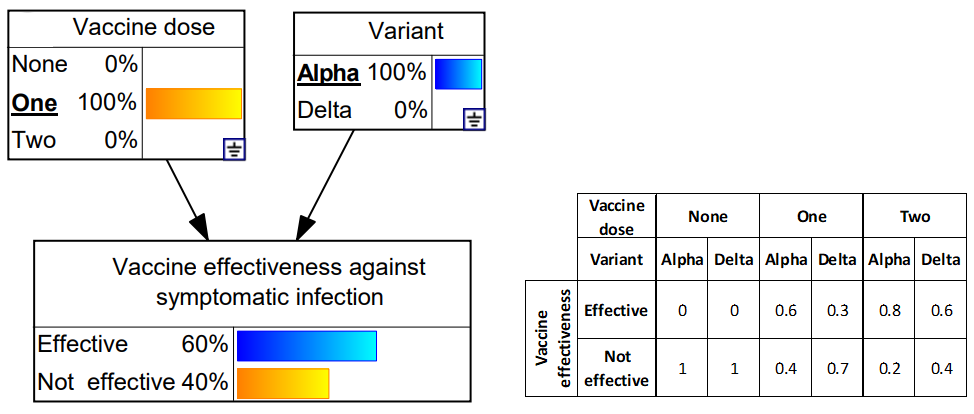}
\caption{An example Bayesian network with a single, dependent child node (Vaccine effectiveness) and two parent nodes (Vaccine doses and variant). Conditional probability table for Vaccine effectiveness is shown on the right.}\label{fig::CoRiCal1}
\end{figure}

Bayesian networks \cite{pearl1988probabilistic} are conditional probability models commonly represented as directed-acyclic graphs, with nodes and links representing variables of interest and the interactions between them. Conditional probabilities for the dependent child-nodes are stored in conditional probability tables (Figure \ref{fig::CoRiCal1}), which determine the probability of a node being in a given state for each possible combination of parent node states. Using Bayes Theorem \cite{pearl1988probabilistic}, the model calculates the probability of a given outcome for any defined scenario. Bayesian networks are widely used in a range of decision support settings including public health \cite{dickson2022bayesian, wu2021bridging}, environmental conservation \cite{camus2022using} and natural resource management \cite{xue2017model}. 

There are several characteristics that make Bayesian networks attractive for an evidence-based COVID-19 risk-benefit calculator. First the conditional probability tables can be populated from different sources, such as data from government reports, results from scientific studies, or recommendations from experts and advisory committees. Second, the probabilistic output means that the model can respond to user-defined scenarios such as “how likely is it that I will get sick” rather than just “will I get sick”. Finally, Bayesian networks are highly interpretable models \cite{uusitalo2007advantages}, as they allow exploration of the effect of different observed values (evidence) on the probability of certain outcomes.

\begin{figure}
\includegraphics[width=1\textwidth]{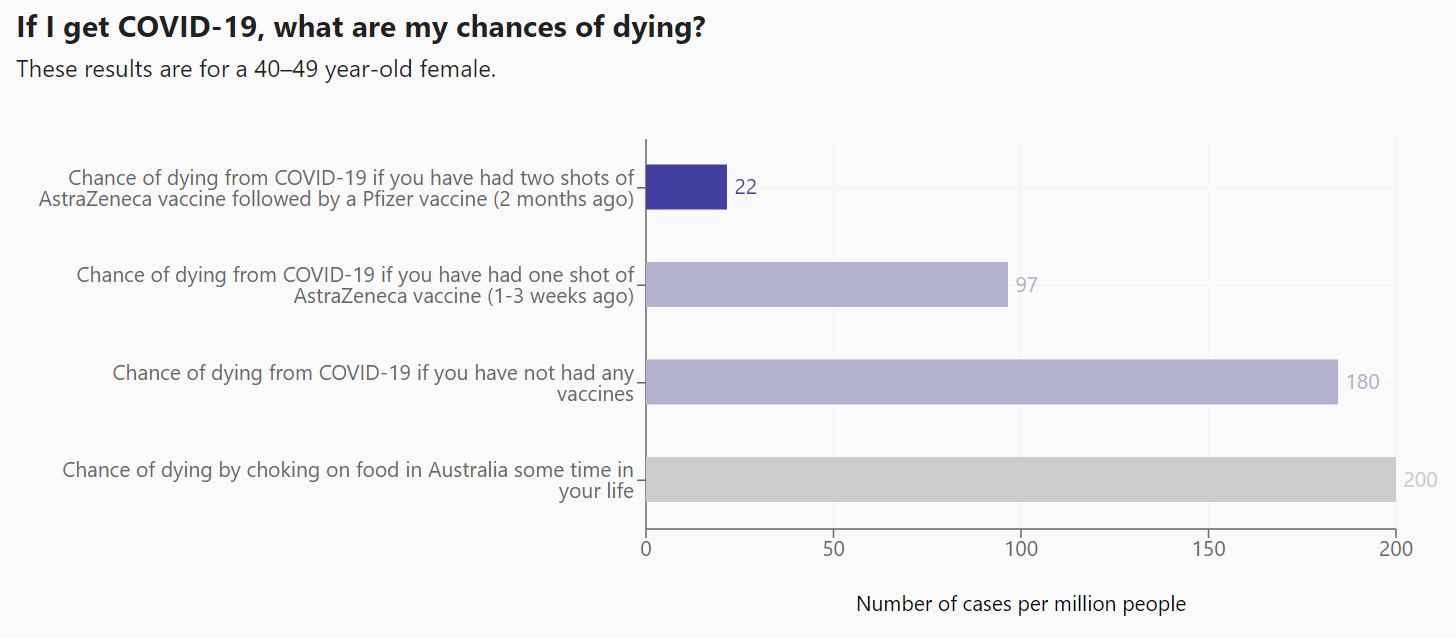}
\caption{An example output from the CoRiCAL COVID-19 risk calculator tool.} \label{fig:CoRiCal2}
\end{figure}

The COVID-19 Risk Calculator (CoRiCAL - \url{https://corical.immunisationcoalition.org.au}) was developed to help the general public, as well as the doctors advising them, weigh-up the risks and benefits of receiving a COVID-19 vaccination.  A Bayesian network model was constructed and parameterised based on the best available evidence from a range of sources that can be used to determine a person’s risk of developing symptomatic COVID-19, dying or other adverse effects from COVID-19, or suffering from adverse effects (including death) from the vaccine itself \cite{mayfield2022designing}.  The model relied on Australian data to represent the context as accurately as possible, however in cases where local data was lacking, international data was used \cite{lau2021risk} \cite{sinclair2022quantifying}. Full model information, along with model code is available via the link \cite{coricalcode}. A web-based interface (Figure \ref{fig:CoRiCal2}) was developed to create a user-friendly tool that considers a person’s age and sex, the brand of the vaccine, how many vaccines they have had already, and the current levels of transmission within the community and displays their chances of an adverse event alongside common relatable risks.  As the pandemic landscape changes, it remains crucial that the evidence for making informed choices on COVID-19 vaccination is made accessible. The model is updated in light of new variants, and as new vaccines become available and recommended (for example booster shots). 

\subsection{ReefCloud: A tool to monitor coral reefs worldwide}


Recent projections estimate that 99\% of the world’s coral reefs will suffer from frequent marine heatwaves under 1.5°C of warming due to climate change \cite{dixon2022future}. Important ecological and socio-economic changes already occur in tropical oceans because of the decline of key corals that support thousands of species \cite{hughes2018spatial}. The latter impacts about one billion people whose income, food supply, coastal protection, and cultural practices depend on coral reef biodiversity. Robust estimation of changes in coral communities at large spatial scales is challenging because there are a lack of observations due to the remoteness of coral reefs and absence of monitoring programs in sea-countries. Also, the fine-scale variability of changes in coral cover result in disparate long-term coral trajectories at reef locations situated only few hundred meters apart \cite{vercelloni2017improved}. For reef managers, these challenges (among others) contribute to slowdown the development of strategies that aim to reduce impacts of climate change on coral reefs.  

A spatio-temporal Bayesian model is developed to estimate the coverage of total coral cover across spatial scales and predict coverage values at unsampled locations. The approach uses outputs from artificial intelligence algorithms trained to classify points on images \cite{gonzalez2020monitoring}. For each Marine Ecoregion Of the World (MEOW, \cite{spalding2007marine}), a set of images $j = 1, 2, \ldots, J$, each composed of $k = 1, 2, \ldots, 50$ elicitation points is used across years of monitoring. Counts, $y_{it}$ for observation $i$ sampled at location $\v s_i$ and time $t$, are modelled using a binomial distribution (with $p$ the probability of positive and $n_i$ the total number of positive cases) and controlled by additional components including the fixed effects of environmental disturbances (cyclones and mass coral bleaching events), sampling nested design (depth, transect, site, reef and monitoring program) modelled as independent and identically distributed Gaussian random effects, and spatio-temporal random effects. 

\begin{figure}[ht]
\includegraphics[width=1\textwidth]{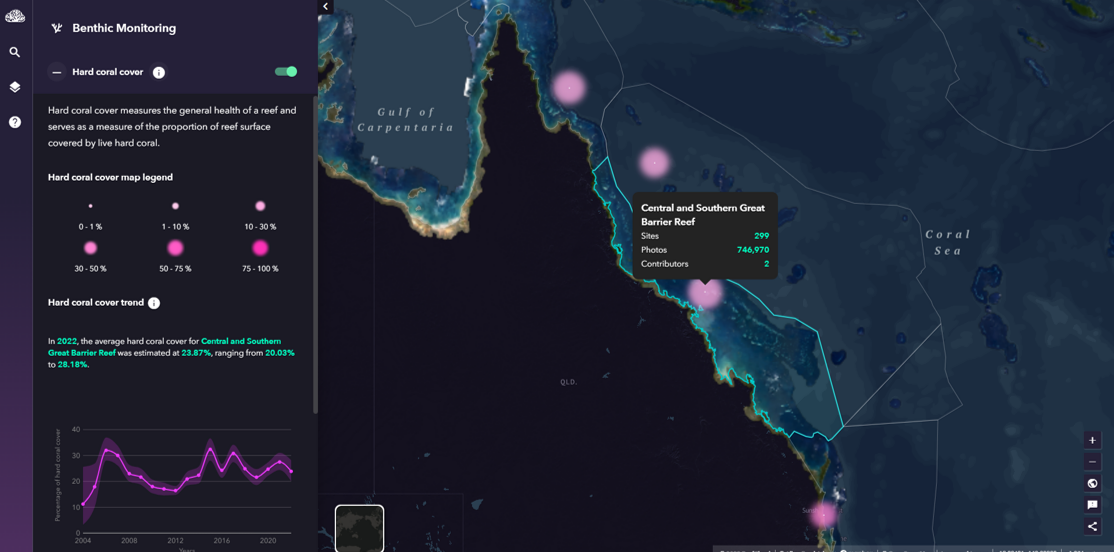}
\caption{An example output from ReefCloud showing temporal trend in coral cover estimating from a Bayesian model for the central and southern parts of the Great Barrier Reef.} \label{ReefCloud}
\end{figure}

The novelty in this model is the incorporation of a spatio-temporal random effects composed of a first-order autoregressive process in time and a Gaussian field that is approximated using a Gaussian Markov random field (GMRF), where the covariance is determined by a Mat{\'e}rn kernel. We employed the GMRF representation as a stochastic partial differential equation, using the method of \cite{Lindgren2015}, implemented in the R package \verb|INLA| \cite{rue2009approximate}. The spatial domain is based on the observed data locations and a buffer with adjacent MEOWs to allow information sharing between units. Spatial predictions are estimated at a grid level of $5\times 5$ km resolution and posterior distributions used to reconstruct coral cover values at coarser spatial scales including MEOWs units and country level. Finally, estimations of coral cover are weighted by the proportion of coral reefs within a MEOW unit following the methodology developed as part of the Global Coral Reef Monitoring Network \cite{souter2020status}. We use the default \verb|INLA| priors for different types of model parameters as discussed in \cite{moraga2019geospatial}. The model is as follows,
\begin{align*}
{y_{it}}| \v \beta,  \v Z, V_{i} &\sim {\text{Binomial}}\bigg(n_i,{\rm logit}^{-1}\bigg(\v \beta^\top \v x_i + r(\v s_i,t) + V_{i}\bigg)\bigg),\\
r({\v s_i,t}) &=  \phi \cdot r(\v s_i,{t-1}) + Z({\v s_i, t}),\\
\v Z(\v s, t) &  \stackrel{\rm ind}{\sim} \mathcal{GP}(\v 0, \m K), \quad t=1,\ldots, T. 
\end{align*} 
The priors for the autoregressive parameter $\phi$ and  independent Gaussian random effects $V_i$ used are the \verb|INLA| defaults.
Research efforts focus on developing new technologies to assess the status of coral reefs in rapid and cost-effective ways through automatic image detection \cite{gonzalez2020monitoring} and learn about impacts of multiple disturbances and management strategies \cite{vercelloni2020forecasting, kennedy2020coral}. ReefCloud is an open-access digital tool that support coral reef monitoring and decision-making by integration of data analyses and reporting (\url{https://reefcloud.ai/}). The online collection of worldwide data provides a unique opportunity to model these data together to 1) increase understanding on the impacts of environmental disturbances and 2) reduce uncertainty when estimating coral trajectories at large spatial scales. The pilot product version is developed using the most extensive monitoring program in the world surveying the Great Barrier Reef, Australia. Machine learning outputs from one million of reef images are used to predict values in coral cover across 3,000 coral reefs from 2004 onward. The ReefCloud online dashboard makes knowledge about reef changes accessible to everyone (Figure \ref{ReefCloud}). The project also educates the reef research community and managers on how Bayesian statistical modelling can help to increase our understanding of the impacts of climate change on coral reefs and supporting decision-making from local to global scales.                  
\subsection{Australian Cancer Atlas}\label{subsect:ACA}
Cancer is the leading cause of disease burden in Australia, which has comprehensive cancer incidence reporting for all cancers except common skin cancers \cite{RN1}. Yet because Australia’s population is heavily concentrated in specific coastal areas, cancer rates are commonly reported only for large regions. Difficulties when using sparse data for smaller areas include the reliability of estimates and the risk of identifying individuals. Yet, detailed small-area analyses have immense power to identify and understand inequities in cancer outcomes.

Using Bayesian hierarchical Poisson models incorporating Leroux priors \cite{RN2}  for spatial smoothing, robust and reliable cancer incidence and 5-year survival estimates were generated across Australian statistical areas level 2 (SA2; 2148 areas). These areas represent communities which interact together and while population sizes vary, the median is around 10,000 people \cite{RN3}. Innovative visualisations helped rapidly identify areas which differed from the national average. Further details on the methods and visualisations are available in \cite{duncan2019development}. Example code for the Bayesian spatial models is available in Sections 7.3 and 9.8.2 of \cite{duncanebook}. 

\begin{figure}
\includegraphics[width=1\textwidth]{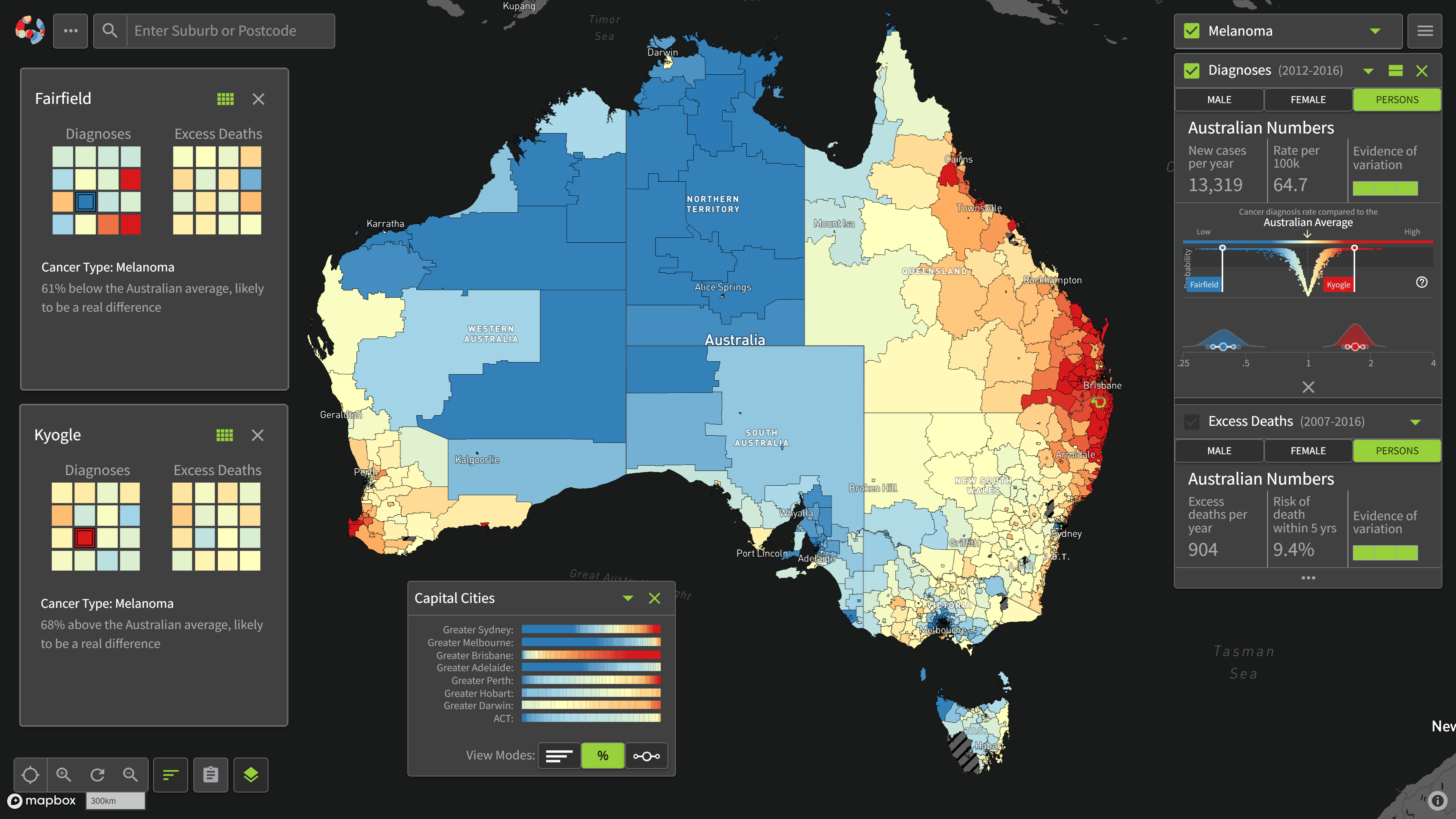}
\caption{An example screenshot of the Australian Cancer Atlas showing melanoma incidence patterns with summary graphs. Red represents high incidence while blue is low in comparison to the national average (pale yellow).} \label{ACA1}
\end{figure}

In September 2018 the Australian Cancer Atlas (\url{atlas.cancer.org.au}) was launched, providing the highest geographic resolution nationwide estimates available (Figure \ref{ACA1}). The website is designed to be interactive and engaging, featuring the ability to download all estimates, export pdfs of specific views, filter regions, rapidly compare different cancer types and rates for two areas, and more! There has been strong uptake and positive feedback, and in 2021 estimates were updated and cancer types expanded.

The Atlas 
has received prestigious spatial industry awards and is currently being replicated internationally.
Australian Cancer Atlas 2.0 is underway, which will examine spatio-temporal patterns, and further include cancer risk factors, some types of cancer treatment and selected cancer clinical/stage patterns. Underpinned by Bayesian methods, the Atlas will continue to provide the methods and visualisations necessary for accurate estimation, interpretation and making decisions.

\section{Conclusion}
This paper has focused on a small number of current opportunities and challenges in the application of the Bayesian paradigm.
Of course, these are not the only issues, but collectively they point to the maturity of current Bayesian practice and the promise of a fully mature Bayesian future.  
As a final thought, we note that many advances in applied Bayesian statistics in recent years are deeply indebted to computational and methodological advances surrounding complex hierarchically-structured models. Modern applied Bayesian statistics thus finds itself at the interface with not only its traditional neighbour mathematics, but also increasingly with the field of computer science. This partnership is one of considerable further promise in the years to come. 
\enlargethispage{20pt}


\dataccess{}

\aucontribute{KB, JM (intelligent data collection); ESF, AP, JV (new data sources); CH, RS (federated analysis); CD, AJ, XW, RS (implicit models); AB, JJB, CD (model transfer); SC, HM, JV (purposeful products); KM (overall); RS (overall editing and coordination).}

\competing{The authors declare that they have no competing interests.}

\funding{AB and JJB acknowledge support from an Australian Research Council Discovery Project (DP200102101); KB was supported by a scholarship under the Australian Research Council Linkage Project (LP180101151),  SC receives salary and research support from an NHMRC Investigator Grant (\#2008313); CD acknowledges support from an Australian Research Council Future Fellowship (FT210100260); JM was supported by an Australian Research Council Discovery (DP200101263) and Linkage Project (LP180101151).
}
\ack{We thank Dr. Jasmine Lee for collecting and providing the 360-degree images in Antarctica.}



\bibliographystyle{rsta} 
\bibliography{bib}

\begin{thebibliography}{100}
\expandafter\ifx\csname urlstyle\endcsname\relax
  \providecommand{\doi}[1]{(doi:\discretionary{}{}{}#1)}\else
  \providecommand{\doi}{(doi:\discretionary{}{}{}\begingroup
  \urlstyle{rm}\Url)}\fi

\bibitem{neyman1928use}
{Neyman, Jerzy and Pearson, Egon S}. 1928 {On the use and interpretation of
  certain test criteria for purposes of statistical inference: Part I}.
\newblock \emph{{Biometrika}} pp. 175--240.

\bibitem{wald1949statistical}
{Wald, Abraham}. 1949 {Statistical decision functions}.
\newblock \emph{{The Annals of Mathematical Statistics}} pp. 165--205.

\bibitem{lindley1972bayesian}
Lindley DV. 1972 \emph{{Bayesian statistics: A review}}.
\newblock SIAM.

\bibitem{muller1995optimal}
M{\"u}ller P, Parmigiani G. 1995 {Optimal design via curve fitting of Monte
  Carlo experiments}.
\newblock \emph{{Journal of the American Statistical Association}} \textbf{90},
  1322--1330.

\bibitem{muller1999simulation}
M{\"u}ller P. 1999 {Simulation-Based Optimal Design} \textbf{6}, 459–--474.

\bibitem{muller2004optimal}
M{\"u}ller P, Sans{\'o} B, De~Iorio M. 2004 {Optimal {B}ayesian design by
  inhomogeneous Markov chain simulation}.
\newblock \emph{Journal of the American Statistical Association} \textbf{99},
  788--798.

\bibitem{amzal2006bayesian}
Amzal B, Bois FY, Parent E, Robert CP. 2006 Bayesian-optimal design via
  interacting particle systems.
\newblock \emph{Journal of the American Statistical association} \textbf{101},
  773--785.

\bibitem{overstall_normal}
Overstall AM, McGree JM, Drovandi CC. 2018 An approach for finding fully
  {B}ayesian optimal designs using normal-based approximations to loss
  functions.
\newblock \emph{Statistics and Computing} \textbf{28}, 343--358.

\bibitem{foster_2019}
Foster A, Jankowiak M, Bingham E, Horsfall P, Teh YW, Rainforth T, Goodman N.
  2019 Variational {B}ayesian optimal experimental design.
\newblock Part of Advances in Neural Information Processing Systems 32 (NeurIPS
  2019)

\bibitem{ace}
Overstall AM, Woods DC. 2017 Bayesian design of experiments using approximate
  coordinate exchange.
\newblock \emph{Technometrics} \textbf{59}, 458--470.

\bibitem{berry2006bayesian}
Berry DA. 2006 Bayesian clinical trials.
\newblock \emph{Nature reviews {Drug} discovery} \textbf{5}, 27--36.

\bibitem{connor_2013}
Connor JT, Elm JJ, Brogliofor KR, {The ESETT and ADAPT-IT investigators}. 2013
  Bayesian adaptive trials for comparative effectiveness research: An example
  in status epilepticus.
\newblock \emph{Journal of Clinical Epidemiology} \textbf{66}, S130--S137.

\bibitem{thorland_2018}
Thorlund K, Haggstrom J, Park JJH, Mills EJ. 2018 Key design considerations for
  adaptive clinical trials: a primer for clinicians.
\newblock \emph{BMJ} \textbf{360}, k698.

\bibitem{kang_2016}
Kang SY, McGree JM, Drovandi C, Mengersen K, Caley J. 2016 Bayesian adaptive
  design: {I}mproving the effectiveness of reef monitoring programs.
\newblock \emph{Ecological Applications} \textbf{26}, 2637--2648.

\bibitem{thilan_2022}
Thilan P, Fisher R, Thompson H, Menendez P, Gilmour J, McGree JM. 2022 Adaptive
  monitoring of coral health at {S}cott {R}eef where data exhibit nonlinear and
  disturbed trends over time.
\newblock \emph{Ecology and Evolution} .Accepted for publication

\bibitem{wagner2020coral}
Wagner D, Friedlander AM, Pyle RL, Brooks CM, Gjerde KM, Wilhelm TA. 2020 Coral
  reefs of the high seas: Hidden biodiversity hotspots in need of protection.
\newblock \emph{Frontiers in Marine Science} \textbf{7}, 1--13.

\bibitem{ltmp2022}
{AIMS}. 2021.
\newblock {Annual Summary Report of Coral Reef Condition 2020/21}.
\newblock
  \url{https://www.aims.gov.au/reef-monitoring/gbr-condition-summary-2020-2021}.

\bibitem{buchhorn2022bayesian}
Buchhorn K, Mengersen K, Santos-Fernandez E, Peterson EE, McGree JM. 2022
  Bayesian design with sampling windows for complex spatial processes.
\newblock \emph{arXiv preprint arXiv:2206.05369} .

\bibitem{bassi2021bayesian}
Bassi A, Berkhof J, de~Jong D, van~de Ven PM. 2021 {Bayesian} adaptive
  decision-theoretic designs for multi-arm multi-stage clinical trials.
\newblock \emph{Statistical Methods in Medical Research} \textbf{30}, 717--730.

\bibitem{giovagnoli2021bayesian}
Giovagnoli A. 2021 The {Bayesian} design of adaptive clinical trials.
\newblock \emph{International Journal of Environmental Research and Public
  Health} \textbf{18}, 530.

\bibitem{kojima2021early}
Kojima M. 2021 Early completion of phase {I} cancer clinical trials with
  {Bayesian} optimal interval design.
\newblock \emph{Statistics in Medicine} \textbf{40}, 3215--3226.

\bibitem{mcgree2022controlled}
McGree J, Hockham C, Kotwal S, Wilcox A, Bassi A, Pollock C, Burrell LM,
  Snelling T, Jha V, Jardine M, \emph{et~al.} 2022 {Controlled evaLuation of
  Angiotensin Receptor Blockers for {COVID}-19 respIraTorY disease (CLARITY):
  Statistical analysis plan for a randomised controlled {Bayesian} adaptive
  sample size trial}.
\newblock \emph{Trials} \textbf{23}, 1--18.

\bibitem{leach2022recursive}
Leach CB, Williams PJ, Eisaguirre JM, Womble JN, Bower MR, Hooten MB. 2022
  Recursive {Bayesian} computation facilitates adaptive optimal design in
  ecological studies.
\newblock \emph{Ecology} \textbf{103}, e03573.

\bibitem{mazumdar2018citizen}
Mazumdar S, Ceccaroni L, Piera J, H{\"o}lker F, Berre A, Arlinghaus R, Bowser
  A. 2018 Citizen science technologies and new opportunities for participation.
\newblock UCL Press.

\bibitem{queiroz2019immersive}
Queiroz ACM, Nascimento AM, Tori R, Silva~Leme MId. 2019 Immersive virtual
  environments and learning assessments.
\newblock In \emph{International Conference on Immersive Learning}, pp.
  172--181. Springer.

\bibitem{fauville2020virtual}
Fauville G, Queiroz ACM, Bailenson JN. 2020 Virtual reality as a promising tool
  to promote climate change awareness.
\newblock \emph{Technology and health} pp. 91--108.

\bibitem{vercelloni2018using}
Vercelloni J, Clifford S, Caley MJ, Pearse AR, Brown R, James A, Christensen B,
  Bednarz T, Anthony K, Gonz{\'a}lez-Rivero M, \emph{et~al.} 2018 Using virtual
  reality to estimate aesthetic values of coral reefs.
\newblock \emph{Royal Society open science} \textbf{5}, 172226.

\bibitem{mengersen2017modelling}
Mengersen K, Peterson EE, Clifford S, Ye N, Kim J, Bednarz T, Brown R, James A,
  Vercelloni J, Pearse AR, \emph{et~al.} 2017 Modelling imperfect presence data
  obtained by citizen science.
\newblock \emph{Environmetrics} \textbf{28}, e2446.

\bibitem{leigh2019using}
Leigh C, Heron G, Wilson E, Gregory T, Clifford S, Holloway J, McBain M,
  Gonzalez F, McGree J, Brown R, \emph{et~al.} 2019 Using virtual reality and
  thermal imagery to improve statistical modelling of vulnerable and protected
  species.
\newblock \emph{PloS one} \textbf{14}, e0217809.

\bibitem{lee2017climate}
Lee JR, Raymond B, Bracegirdle TJ, Chad{\`e}s I, Fuller RA, Shaw JD, Terauds A.
  2017 Climate change drives expansion of {A}ntarctic ice-free habitat.
\newblock \emph{Nature} \textbf{547}, 49--54.

\bibitem{parties1960protocol}
Parties ATC. 1960.
\newblock Protocol on {E}nvironmental {P}rotection to the {A}ntarctic {T}reaty.

\bibitem{vercelloni2021connecting}
Vercelloni J, Peppinck J, Santos-Fernandez E, McBain M, Heron G, Dodgen T,
  Peterson EE, Mengersen K. 2021 Connecting virtual reality and ecology: a new
  tool to run seamless immersive experiments in {R}.
\newblock \emph{PeerJ Computer Science} \textbf{7}, e544.

\bibitem{zooniverse}
{Zooniverse}. 2022.
\newblock {Zooniverse}.
\newblock \url{https://www.zooniverse.org}.
\newblock Accessed:2022-09-23

\bibitem{prudic2017ebutterfly}
Prudic KL, McFarland KP, Oliver JC, Hutchinson RA, Long EC, Kerr JT,
  Larriv{\'e}e M. 2017 {eButterfly: leveraging massive online citizen science
  for butterfly conservation}.
\newblock \emph{Insects} \textbf{8}, 53.

\bibitem{sullivan2009ebird}
Sullivan BL, Wood CL, Iliff MJ, Bonney RE, Fink D, Kelling S. 2009 {eBird: A
  citizen-based bird observation network in the biological sciences}.
\newblock \emph{Biological Conservation} \textbf{142}, 2282--2292.

\bibitem{nugent2018inaturalist}
Nugent J. 2018 inaturalist.
\newblock \emph{Science Scope} \textbf{41}, 12--13.

\bibitem{van2013occupancy}
van Strien AJ, Termaat T, Kalkman V, Prins M, De~Knijf G, Gourmand AL, Houard
  X, Nelson B, Plate C, Prentice S, \emph{et~al.} 2013 Occupancy modelling as a
  new approach to assess supranational trends using opportunistic data: a pilot
  study for the damselfly calopteryx splendens.
\newblock \emph{Biodiversity and Conservation} \textbf{22}, 673--686.

\bibitem{dwyer2016using}
Dwyer RG, Carpenter-Bundhoo L, Franklin CE, Campbell HA. 2016 Using
  citizen-collected wildlife sightings to predict traffic strike hot spots for
  threatened species: a case study on the southern cassowary.
\newblock \emph{Journal of Applied Ecology} \textbf{53}, 973--982.

\bibitem{strebel2014studying}
Strebel N, K{\'e}ry M, Schaub M, Schmid H. 2014 Studying phenology by flexible
  modelling of seasonal detectability peaks.
\newblock \emph{Methods in Ecology and Evolution} \textbf{5}, 483--490.

\bibitem{santos2021MEE}
Santos-Fern{\'a}ndez E, Mengersen K. 2021 Understanding the reliability of
  citizen science observational data using item response models.
\newblock \emph{Methods in Ecology and Evolution} \textbf{12}, 1533--1548.

\bibitem{freitag2016strategies}
Freitag A, Meyer R, Whiteman L. 2016 Strategies employed by citizen science
  programs to increase the credibility of their data.
\newblock \emph{Citizen Science: Theory and Practice} \textbf{1}.

\bibitem{santos2020}
Santos-Fernandez E, Peterson EE, Vercelloni J, Rushworth E, Mengersen K. 2021
  Correcting misclassification errors in crowdsourced ecological data: A
  {B}ayesian perspective.
\newblock \emph{Journal of the Royal Statistical Society: Series C}
  \textbf{70}, 147--173.

\bibitem{baker2004item}
Baker FB, Kim SH. 2004 \emph{{Item response theory: Parameter estimation
  techniques}}.
\newblock CRC Press.

\bibitem{hakuna}
{EdgarSantos-Fernandez}. 2021.
\newblock {hakuna}.
\newblock \url{https://github.com/EdgarSantos-Fernandez/hakuna}.

\bibitem{peterson2020monitoring}
Peterson EE, Santos-Fern{\'a}ndez E, Chen C, Clifford S, Vercelloni J, Pearse
  A, Brown R, Christensen B, James A, Anthony K, \emph{et~al.} 2020 Monitoring
  through many eyes: Integrating disparate datasets to improve monitoring of
  the great barrier reef.
\newblock \emph{Environmental Modelling \& Software} \textbf{124}, 104557.

\bibitem{mcmahan2017communication}
McMahan HB, Moore E, Ramage D, Hampson S, y~Arcas BA. 2017
  Communication-efficient learning of deep networks from decentralized data.

\bibitem{li2020federated}
Li T, Sahu AK, Zaheer M, Sanjabi M, Talwalkar A, Smith V. 2020 Federated
  optimization in heterogeneous networks.
\newblock \emph{Proceedings of Machine Learning and Systems} \textbf{2},
  429--450.

\bibitem{wang2020tackling}
Wang J, Liu Q, Liang H, Joshi G, Poor HV. 2020 Tackling the objective
  inconsistency problem in heterogeneous federated optimization.
\newblock \emph{Advances in neural information processing systems} \textbf{33},
  7611--7623.

\bibitem{yurochkin2019bayesian}
Yurochkin M, Agarwal M, Ghosh S, Greenewald K, Hoang N, Khazaeni Y. 2019
  {B}ayesian nonparametric federated learning of neural networks.
\newblock In \emph{International Conference on Machine Learning}, pp.
  7252--7261. PMLR.

\bibitem{thibaux2007hierarchical}
Thibaux R, Jordan MI. 2007 Hierarchical beta processes and the {I}ndian buffet
  process.
\newblock In \emph{Conference on Artificial intelligence and statistics}, pp.
  564--571. PMLR.

\bibitem{deist2020distributed}
Deist TM, Dankers FJ, Ojha P, Marshall MS, Janssen T, Faivre-Finn C, Masciocchi
  C, Valentini V, Wang J, Chen J, \emph{et~al.} 2020 {Distributed learning on
  20 000+ lung cancer patients--The Personal Health Train}.
\newblock \emph{Radiotherapy and Oncology} \textbf{144}, 189--200.

\bibitem{geleijnse2020prognostic}
Geleijnse G, Chiang RCJ, Sieswerda M, Schuurman M, Lee K, van Soest J, Dekker
  A, Lee WC, Verbeek XA. 2020 Prognostic factors for survival in patients with
  oral cavity cancer: a comparison of the {N}etherlands and {T}aiwan using
  privacy-preserving federated analyses.
\newblock \emph{BMC Medical Informatics and Decision Making} .

\bibitem{cellamare2022federated}
Cellamare M, van Gestel AJ, Alradhi H, Martin F, Moncada-Torres A. 2022 A
  federated generalized linear model for privacy-preserving analysis.
\newblock \emph{Algorithms} .

\bibitem{fienberg2006secure}
Fienberg SE, Fulp WJ, Slavkovic AB, Wrobel TA. 2006 “{S}ecure” log-linear
  and logistic regression analysis of distributed databases.
\newblock In \emph{International Conference on Privacy in Statistical
  Databases}, pp. 277--290. Springer.

\bibitem{slavkovic2007secure}
Slavkovic AB, Nardi Y, Tibbits MM. 2007 "{S}ecure" logistic regression of
  horizontally and vertically partitioned distributed databases.
\newblock In \emph{Seventh IEEE International Conference on Data Mining
  Workshops (ICDMW 2007)}, pp. 723--728. IEEE.

\bibitem{shi2016secure}
Shi H, Jiang C, Dai W, Jiang X, Tang Y, Ohno-Machado L, Wang S. 2016 {S}ecure
  {M}ulti-p{A}rty {C}omputation {G}rid {LO}gistic {RE}gression ({SMAC-GLORE}).
\newblock \emph{BMC Medical Informatics and Decision Making} \textbf{16},
  175--187.

\bibitem{li2016vertical}
Li Y, Jiang X, Wang S, Xiong H, Ohno-Machado L. 2016 Vertical grid logistic
  regression ({VERTIGO}).
\newblock \emph{Journal of the American Medical Informatics Association}
  \textbf{23}, 570--579.

\bibitem{kamphorst2022accurate}
Kamphorst B, Rooijakkers T, Veugen T, Cellamare M, Knoors D. 2022 {Accurate
  training of the Cox proportional hazards model on vertically-partitioned data
  while preserving privacy}.
\newblock \emph{BMC Medical Informatics and Decision Making} \textbf{22},
  1--18.

\bibitem{cramer2015secure}
Cramer R, Damg{\aa}rd IB, \emph{et~al.} 2015 \emph{Secure multiparty
  computation}.
\newblock Cambridge University Press.

\bibitem{minka2003comparison}
Minka TP. 2003 A comparison of numerical optimizers for logistic regression.
\newblock \emph{Unpublished draft} pp. 1--18.

\bibitem{moncada2020vantage6}
Moncada-Torres A, Martin F, Sieswerda M, Van~Soest J, Geleijnse G. 2020
  {VANTAGE6}: an open source pri{VA}cy preservi{N}g federa{T}ed le{A}rnin{G}
  infrastructur{E} for {S}ecure {I}nsight e{X}change.
\newblock In \emph{AMIA Annual Symposium Proceedings}, volume 2020, p. 870.
  American Medical Informatics Association.

\bibitem{wang2014parallelizing}
Wang X, Dunson DB. 2014 Parallelizing {MCMC} via {W}eierstrass sampler.
\newblock \emph{arXiv preprint arXiv:1312.4605} .

\bibitem{neiswanger2014asymptotically}
Neiswanger W, Wang C, Xing EP. 2014 Asymptotically exact, embarrassingly
  parallel {MCMC}.
\newblock In \emph{Proceedings of the Thirtieth Conference on Uncertainty in
  Artificial Intelligence}, UAI'14, p. 623–632. Arlington, Virginia, USA:
  AUAI Press.

\bibitem{scott2016bayes}
Scott SL, Blocker AW, Bonassi FV, Chipman HA, George EI, McCulloch RE. 2016
  Bayes and big data: The consensus {M}onte carlo algorithm.
\newblock \emph{International Journal of Management Science and Engineering
  Management} \textbf{11}, 78--88.

\bibitem{jordan2019communication}
Jordan MI, Lee JD, Yang Y. 2019 {Communication-efficient distributed
  statistical inference}.
\newblock \emph{Journal of the American Statistical Association} \textbf{114},
  668--681.

\bibitem{plassier2021dg}
Plassier V, Vono M, Durmus A, Moulines E. 2021 {DG-LMC}: a turn-key and
  scalable synchronous distributed {MCMC} algorithm via {L}angevin {M}onte
  {C}arlo within {G}ibbs.
\newblock In \emph{International Conference on Machine Learning}, pp.
  8577--8587. PMLR.

\bibitem{el2021federated}
El~Mekkaoui K, Mesquita D, Blomstedt P, Kaski S. 2021 Federated stochastic
  gradient {L}angevin dynamics.
\newblock In \emph{Uncertainty in Artificial Intelligence}, pp. 1703--1712.
  PMLR.

\bibitem{de2020overview}
De~Cristofaro E. 2020 An overview of privacy in machine learning.
\newblock \emph{arXiv preprint arXiv:2005.08679} .

\bibitem{heikkila2019differentially}
Heikkil{\"a} M, J{\"a}lk{\"o} J, Dikmen O, Honkela A. 2019 Differentially
  private {M}arkov chain {M}onte {C}arlo.
\newblock \emph{Advances in Neural Information Processing Systems} \textbf{32}.

\bibitem{bohensky2010data}
Bohensky MA, Jolley D, Sundararajan V, Evans S, Pilcher DV, Scott I, Brand CA.
  2010 Data linkage: a powerful research tool with potential problems.
\newblock \emph{BMC health services research} \textbf{10}, 1--7.

\bibitem{harron2017challenges}
Harron K, Dibben C, Boyd J, Hjern A, Azimaee M, Barreto ML, Goldstein H. 2017
  Challenges in administrative data linkage for research.
\newblock \emph{Big data \& society} \textbf{4}, 2053951717745678.

\bibitem{gelman2006data}
Gelman A, Hill J. 2006 \emph{Data analysis using regression and
  multilevel/hierarchical models}.
\newblock Cambridge university press.

\bibitem{besag1974spatial}
Besag J. 1974 Spatial interaction and the statistical analysis of lattice
  systems.
\newblock \emph{Journal of the Royal Statistical Society: Series B
  (Methodological)} \textbf{36}, 192--225.

\bibitem{besag1991bayesian}
Besag J, York J, Molli{\'e} A. 1991 {B}ayesian image restoration, with two
  applications in spatial statistics.
\newblock \emph{Annals of the institute of statistical mathematics}
  \textbf{43}, 1--20.

\bibitem{leroux2000estimation}
Leroux BG, Lei X, Breslow N. 2000 Estimation of disease rates in small areas: a
  new mixed model for spatial dependence.
\newblock In \emph{Statistical models in epidemiology, the environment, and
  clinical trials}, pp. 179--191. Springer.

\bibitem{Sisson2018}
Sisson SA, Fan Y, Beaumont M. 2018 \emph{Handbook of Approximate {B}ayesian
  Computation}.
\newblock Chapman and Hall/CRC, 1st edition.

\bibitem{Beaumont2002}
Beaumont MA, Zhang W, Balding DJ. 2002 {Approximate {B}ayesian computation in
  population genetics}.
\newblock \emph{Genetics} \textbf{162}, 2025--2035.

\bibitem{beven1992future}
Beven K, Binley A. 1992 The future of distributed models: model calibration and
  uncertainty prediction.
\newblock \emph{Hydrological processes} \textbf{6}, 279--298.

\bibitem{beven2014glue}
Beven K, Binley A. 2014 Glue: 20 years on.
\newblock \emph{Hydrological processes} \textbf{28}, 5897--5918.

\bibitem{nott2012generalized}
Nott DJ, Marshall L, Brown J. 2012 Generalized likelihood uncertainty
  estimation (glue) and approximate {B}ayesian computation: What's the
  connection?
\newblock \emph{Water Resources Research} \textbf{48}.

\bibitem{prangle2018summary}
Prangle D. 2018 Summary statistics.
\newblock In \emph{Handbook of approximate {B}ayesian computation}, pp.
  125--152. Chapman and Hall/CRC.

\bibitem{drovandi2022comparison}
Drovandi C, Frazier DT. 2022 A comparison of likelihood-free methods with and
  without summary statistics.
\newblock \emph{Statistics and Computing} \textbf{32}, 1--23.

\bibitem{Sisson2018a}
Sisson S, Fan Y. 2018 \emph{Handbook of approximate {B}ayesian computation},
  chapter {ABC} samplers, pp. 87--123.
\newblock Chapman and Hall/CRC.

\bibitem{frazier+mrr18}
Frazier DT, Martin GM, Robert CP, Rousseau J. 2018 Asymptotic properties of
  approximate {B}ayesian computation.
\newblock \emph{Biometrika} \textbf{105}, 593--607.

\bibitem{Price2018}
Price LF, Drovandi CC, Lee A, Nott DJ. 2018 Bayesian synthetic likelihood.
\newblock \emph{Journal of Computational and Graphical Statistics} \textbf{27},
  1--11.

\bibitem{frazier2020BSLasymp}
Frazier D, Nott DJ, Drovandi C, Kohn R. 2022 Bayesian inference using synthetic
  likelihood: asymptotics and adjustments.
\newblock \emph{Journal of the American Statistical Association} .

\bibitem{carr2021estimating}
Carr MJ, Simpson MJ, Drovandi C. 2021 Estimating parameters of a stochastic
  cell invasion model with fluorescent cell cycle labelling using approximate
  {B}ayesian computation.
\newblock \emph{Journal of the Royal Society Interface} \textbf{18}, 20210362.

\bibitem{Drovandi2011b}
Drovandi CC, Pettitt AN. 2011 Estimation of parameters for macroparasite
  population evolution using approximate {B}ayesian computation.
\newblock \emph{Biometrics} \textbf{67}, 225--233.

\bibitem{papamakarios2019sequential}
Papamakarios G, Sterratt D, Murray I. 2019 Sequential neural likelihood: Fast
  likelihood-free inference with autoregressive flows.
\newblock In \emph{The 22nd International Conference on Artificial Intelligence
  and Statistics}, pp. 837--848. PMLR.

\bibitem{thomas2022likelihood}
Thomas O, Dutta R, Corander J, Kaski S, Gutmann MU. 2022 Likelihood-free
  inference by ratio estimation.
\newblock \emph{Bayesian Analysis} \textbf{17}, 1--31.

\bibitem{lueckmann2017flexible}
Lueckmann JM, Goncalves PJ, Bassetto G, {\"O}cal K, Nonnenmacher M, Macke JH.
  2017 Flexible statistical inference for mechanistic models of neural
  dynamics.
\newblock \emph{Advances in {N}eural {I}nformation {P}rocessing {S}ystems}
  \textbf{30}.

\bibitem{wang2015simulating}
Wang Z, Butner JD, Kerketta R, Cristini V, Deisboeck TS. 2015 Simulating cancer
  growth with multiscale agent-based modeling.
\newblock In \emph{Seminars in {C}ancer {B}iology}, volume~30, pp. 70--78.
  Elsevier.

\bibitem{metzcar2019review}
Metzcar J, Wang Y, Heiland R, Macklin P. 2019 A review of cell-based
  computational modeling in cancer biology.
\newblock \emph{JCO {C}linical {C}ancer {I}nformatics} \textbf{2}, 1--13.

\bibitem{macnamara2021biomechanical}
Macnamara CK. 2021 Biomechanical modelling of cancer: Agent-based force-based
  models of solid tumours within the context of the tumour microenvironment.
\newblock \emph{Computational and Systems Oncology} \textbf{1}, e1018.

\bibitem{cess2022multiscale}
Cess CG, Finley SD. 2022 Multiscale modeling of tumor adaption and invasion
  following anti-angiogenic therapy.
\newblock \emph{Computational and Systems Oncology} \textbf{2}, e1032.

\bibitem{norton2019multiscale}
Norton KA, Gong C, Jamalian S, Popel AS. 2019 Multiscale agent-based and hybrid
  modeling of the tumor immune microenvironment.
\newblock \emph{{P}rocesses} \textbf{7}, 37.

\bibitem{cess2020multi}
Cess CG, Finley SD. 2020 {Multi-scale modeling of macrophage—T cell
  interactions within the tumor microenvironment}.
\newblock \emph{PLoS {C}omputational {B}iology} \textbf{16}, e1008519.

\bibitem{jenner2022agent}
Jenner AL, Smalley M, Goldman D, Goins WF, Cobbs CS, Puchalski RB, Chiocca EA,
  Lawler S, Macklin P, Goldman A, \emph{et~al.} 2022 Agent-based computational
  modeling of glioblastoma predicts that stromal density is central to
  oncolytic virus efficacy.
\newblock \emph{iScience} \textbf{25}.

\bibitem{klowss2022stochastic}
Klowss JJ, Browning AP, Murphy RJ, Carr EJ, Plank MJ, Gunasingh G, Haass NK,
  Simpson MJ. 2022 A stochastic mathematical model of 4{D} tumour spheroids
  with real-time fluorescent cell cycle labelling.
\newblock \emph{Journal of the Royal Society Interface} \textbf{19}, 20210903.

\bibitem{gallaher2020cells}
Gallaher JA, Massey SC, Hawkins-Daarud A, Noticewala SS, Rockne RC, Johnston
  SK, Gonzalez-Cuyar L, Juliano J, Gil O, Swanson KR, \emph{et~al.} 2020 {From
  cells to tissue: How cell scale heterogeneity impacts glioblastoma growth and
  treatment response}.
\newblock \emph{PLoS {C}omputational {B}iology} \textbf{16}, e1007672.

\bibitem{jenner2022examining}
Jenner A, Kelly W, Dallaston M, Araujo R, Parfitt I, Steinitz D, Pooladvand P,
  Kim PS, Wade SJ, Vine KL. 2022 Examining the efficacy of localised
  gemcitabine therapy for the treatment of pancreatic cancer using a hybrid
  agent-based model.
\newblock \emph{bioRxiv} .

\bibitem{jenner2020enhancing}
Jenner AL, Frascoli F, Coster AC, Kim PS. 2020 Enhancing oncolytic virotherapy:
  Observations from a {V}oronoi {C}ell-based model.
\newblock \emph{Journal of Theoretical Biology} \textbf{485}, 110052.

\bibitem{kim2011active}
Kim PH, Sohn JH, Choi JW, Jung Y, Kim SW, Haam S, Yun CO. 2011 Active targeting
  and safety profile of peg-modified adenovirus conjugated with herceptin.
\newblock \emph{Biomaterials} \textbf{32}, 2314--2326.

\bibitem{wang2022calibration}
Wang X, Jenner AL, Salomone R, Drovandi C. 2022 Calibration of a {V}oronoi
  cell-based model for tumour growth using approximate {B}ayesian computation.
\newblock \emph{bioRxiv} .

\bibitem{tejero2020sbi}
Tejero-Cantero A, Boelts J, Deistler M, Lueckmann JM, Durkan C, Gon{\c{c}}alves
  PJ, Greenberg DS, Macke JH. 2020 {SBI}--a toolkit for simulation-based
  inference.
\newblock \emph{arXiv preprint arXiv:2007.09114} .

\bibitem{lueckmann2021benchmarking}
Lueckmann JM, Boelts J, Greenberg D, Goncalves P, Macke J. 2021 Benchmarking
  simulation-based inference.
\newblock In \emph{International Conference on Artificial Intelligence and
  Statistics}, pp. 343--351. PMLR.

\bibitem{ABCSNLgit}
{john-wang1015}. 2022.
\newblock {ABCandSNL}.
\newblock \url{https://github.com/john-wang1015/ABCandSNL}.

\bibitem{zhuang2020comprehensive}
Zhuang F, Qi Z, Duan K, Xi D, Zhu Y, Zhu H, Xiong H, He Q. 2020 {A
  Comprehensive Survey on Transfer Learning}.
\newblock \emph{Proceedings of the IEEE} \textbf{109}, 43--76.

\bibitem{agarwal2021transfer}
Agarwal N, Sondhi A, Chopra K, Singh G. 2021 {Transfer learning: Survey and
  classification}.
\newblock \emph{Smart innovations in communication and computational sciences}
  pp. 145--155.

\bibitem{YE202229}
Ye K, Han Z, Duan Y, Bai T. 2022 {Normalized power prior {B}ayesian analysis}.
\newblock \emph{Journal of Statistical Planning and Inference} \textbf{216},
  29--50.

\bibitem{pawel2022normalized}
Pawel S, Aust F, Held L, Wagenmakers EJ. 2022 {Normalized power priors always
  discount historical data}.
\newblock \emph{arXiv preprint arXiv:2206.04379} .

\bibitem{ibrahim2015power}
Ibrahim JG, Chen MH, Gwon Y, Chen F. 2015 {The power prior: theory and
  applications}.
\newblock \emph{Statistics in medicine} \textbf{34}, 3724--3749.

\bibitem{han2022study}
Han Z, Ye K, Wang M. 2022 {A Study on the Power Parameter in Power Prior
  {B}ayesian Analysis}.
\newblock \emph{The American Statistician} pp. 1--8.

\bibitem{bennett2021novel}
Bennett M, White S, Best N, Mander A. 2021 {A novel equivalence probability
  weighted power prior for using historical control data in an adaptive
  clinical trial design: A comparison to standard methods}.
\newblock \emph{Pharmaceutical statistics} \textbf{20}, 462--484.

\bibitem{app10020559}
Chouhan V, Singh SK, Khamparia A, Gupta D, Tiwari P, Moreira C, Damasevicius R,
  de~Albuquerque VHC. 2020 {A Novel Transfer Learning Based Approach for
  Pneumonia Detection in Chest X-ray Images}.
\newblock \emph{Applied Sciences} \textbf{10}.

\bibitem{gupta2022boosting}
Gupta S, Bi J, Liu Y, Wildani A. 2022 {Boosting For Regression Transfer via
  Importance Sampling} .

\bibitem{tang2020improving}
Tang D, Yang X, Wang X. 2020 {Improving the transferability of the crash
  prediction model using the TrAdaBoost. R2 algorithm}.
\newblock \emph{Accident Analysis \& Prevention} \textbf{141}, 105551.

\bibitem{solomatine2004adaboost}
Solomatine DP, Shrestha DL. 2004 {AdaBoost. RT: a boosting algorithm for
  regression problems}.
\newblock In \emph{2004 IEEE International Joint Conference on Neural Networks
  (IEEE Cat. No. 04CH37541)}, volume~2, pp. 1163--1168. IEEE.

\bibitem{li2021transfer}
Li S, Ren Z, Sabatti C, Sesia M. 2021 {Transfer learning in genome-wide
  association studies with knockoffs}.
\newblock \emph{arXiv preprint arXiv:2108.08813} .

\bibitem{maity2021linear}
Maity S, Dutta D, Terhorst J, Sun Y, Banerjee M. 2021 {A linear adjustment
  based approach to posterior drift in transfer learning}.
\newblock \emph{arXiv preprint arXiv:2111.10841} .

\bibitem{reeve2021adaptive}
Reeve HW, Cannings TI, Samworth RJ. 2021 {Adaptive transfer learning}.
\newblock \emph{The Annals of Statistics} \textbf{49}, 3618--3649.

\bibitem{guo2017towards}
Guo S, Heinke R, St{\"o}ckel S, R{\"o}sch P, Bocklitz T, Popp J. 2017 {Towards
  an improvement of model transferability for Raman spectroscopy in biological
  applications}.
\newblock \emph{Vibrational Spectroscopy} \textbf{91}, 111--118.

\bibitem{hector2022turning}
Hector EC, Martin R. 2022 {Turning the information-sharing dial: efficient
  inference from different data sources}.
\newblock \emph{arXiv preprint arXiv:2207.08886} .

\bibitem{da2019fast}
Da B, Ong YS, Gupta A, Feng L, Liu H. 2019 {Fast transfer Gaussian process
  regression with large-scale sources}.
\newblock \emph{Knowledge-Based Systems} \textbf{165}, 208--218.

\bibitem{wei2022transfer}
Wei P, Vo TV, Qu X, Ong YS, Ma Z. 2022 {Transfer Kernel Learning for
  Multi-Source Transfer Gaussian Process Regression}.
\newblock \emph{IEEE Transactions on Pattern Analysis and Machine Intelligence}
  .

\bibitem{cao2010adaptive}
Cao B, Pan SJ, Zhang Y, Yeung DY, Yang Q. 2010 Adaptive transfer learning.
\newblock In \emph{proceedings of the AAAI Conference on Artificial
  Intelligence}, volume~24, pp. 407--412.

\bibitem{tgpgit}
{Xiao-dong-Wang}. 2021.
\newblock {Transfer-GP}.
\newblock \url{https://github.com/Xiao-dong-Wang/Transfer-GP}.

\bibitem{gov}
{Australian Government Department of Health and Aged Care}. 2022.
\newblock {First COVID-19 vaccinations in Australia 2021.}
\newblock \url{
  https://www.health.gov.au/news/first-covid-19-vaccinations-in-australia}

\bibitem{greinacher2021thrombotic}
Greinacher A, Thiele T, Warkentin TE, Weisser K, Kyrle PA, Eichinger S. 2021
  Thrombotic thrombocytopenia after {C}h{A}d{O}x1 n{C}ov-19 vaccination.
\newblock \emph{New England Journal of Medicine} \textbf{384}, 2092--2101.

\bibitem{marshall2021symptomatic}
Marshall M, Ferguson ID, Lewis P, Jaggi P, Gagliardo C, Collins JS, Shaughnessy
  R, Caron R, Fuss C, Corbin KJE, \emph{et~al.} 2021 Symptomatic acute
  myocarditis in 7 adolescents after {P}fizer-{B}io{NT}ech {COVID}-19
  vaccination.
\newblock \emph{Pediatrics} \textbf{148}.

\bibitem{leask2021communicating}
Leask J, Carlson SJ, Attwell K, Clark KK, Kaufman J, Hughes C, Frawley J,
  Cashman P, Seal H, Wiley K, \emph{et~al.} 2021 Communicating with patients
  and the public about {COVID}-19 vaccine safety: recommendations from the
  collaboration on social science and immunisation.
\newblock \emph{Med J Aust} \textbf{215}, 9--12.

\bibitem{sheikh2021sars}
Sheikh A, McMenamin J, Taylor B, Robertson C. 2021 {SARS-CoV}-2 {Delta} {VOC}
  in {S}cotland: demographics, risk of hospital admission, and vaccine
  effectiveness.
\newblock \emph{The Lancet} \textbf{397}, 2461--2462.

\bibitem{zheng2022real}
Zheng C, Shao W, Chen X, Zhang B, Wang G, Zhang W. 2022 Real-world
  effectiveness of {COVID}-19 vaccines: a literature review and meta-analysis.
\newblock \emph{International Journal of Infectious Diseases} \textbf{114},
  252--260.

\bibitem{pearl1988probabilistic}
Pearl J. 1988 \emph{Probabilistic reasoning in intelligent systems: networks of
  plausible inference}.
\newblock Morgan kaufmann.

\bibitem{dickson2022bayesian}
Dickson BF, Masson JJ, Mayfield HJ, Aye KS, Htwe KM, Roineau M, Andreosso A,
  Ryan S, Becker L, Douglass J, \emph{et~al.} 2022 Bayesian network analysis of
  lymphatic filariasis serology from myanmar shows benefit of adding antibody
  testing to post-mda surveillance.
\newblock \emph{Tropical Medicine and Infectious Disease} \textbf{7}, 113.

\bibitem{wu2021bridging}
Wu Y, Foley D, Ramsay J, Woodberry O, Mascaro S, Nicholson AE, Snelling T. 2021
  Bridging the gaps in test interpretation of {SARS}-{C}o{V}-2 through
  {B}ayesian network modelling.
\newblock \emph{Epidemiology \& Infection} \textbf{149}.

\bibitem{camus2022using}
Camus EB, Rhodes JR, McAlpine CA, Lunney D, Callaghan J, Goldingay R, Brace A,
  Hall M, Hetherington SB, Hopkins M, \emph{et~al.} 2022 Using expert
  elicitation to identify effective combinations of management actions for
  koala conservation in different regional landscapes.
\newblock \emph{Wildlife Research} .

\bibitem{xue2017model}
Xue J, Gui D, Lei J, Zeng F, Mao D, Zhang Z. 2017 Model development of a
  participatory {B}ayesian network for coupling ecosystem services into
  integrated water resources management.
\newblock \emph{Journal of Hydrology} \textbf{554}, 50--65.

\bibitem{uusitalo2007advantages}
Uusitalo L. 2007 Advantages and challenges of {B}ayesian networks in
  environmental modelling.
\newblock \emph{Ecological modelling} \textbf{203}, 312--318.

\bibitem{mayfield2022designing}
Mayfield HJ, Lau CL, Sinclair JE, Brown SJ, Baird A, Litt J, Vuorinen A, Short
  KR, Waller M, Mengersen K. 2022 Designing an evidence-based {B}ayesian
  network for estimating the risk versus benefits of {AstraZeneca} {COVID}-19
  vaccine.
\newblock \emph{Vaccine} \textbf{40}, 3072--3084.

\bibitem{lau2021risk}
Lau CL, Mayfield HJ, Sinclair JE, Brown SJ, Waller M, Enjeti AK, Baird A, Short
  KR, Mengersen K, Litt J. 2021 Risk-benefit analysis of the {AstraZeneca}
  {COVID}-19 vaccine in {A}ustralia using a {B}ayesian network modelling
  framework.
\newblock \emph{Vaccine} \textbf{39}, 7429--7440.

\bibitem{sinclair2022quantifying}
Sinclair JE, Mayfield HJ, Short KR, Brown SJ, Puranik R, Mengersen K, Litt JC,
  Lau CL. 2022 {A {B}ayesian network analysis quantifying risks versus benefits
  of the Pfizer COVID-19 vaccine in {A}ustralia}.
\newblock \emph{npj Vaccines} \textbf{7}, 1--11.

\bibitem{coricalcode}
{BayesFusion Interactive Model Repository: CoRiCal AstraZeneca Model}.
\newblock
  \url{https://repo.bayesfusion.com/network/permalink?net=Small+BNs\%2FCoRiCalAZ.xdsl}

\bibitem{dixon2022future}
Dixon AM, Forster PM, Heron SF, Stoner AM, Beger M. 2022 Future loss of
  local-scale thermal refugia in coral reef ecosystems.
\newblock \emph{Plos Climate} \textbf{1}, e0000004.

\bibitem{hughes2018spatial}
Hughes TP, Anderson KD, Connolly SR, Heron SF, Kerry JT, Lough JM, Baird AH,
  Baum JK, Berumen ML, Bridge TC, \emph{et~al.} 2018 Spatial and temporal
  patterns of mass bleaching of corals in the anthropocene.
\newblock \emph{Science} \textbf{359}, 80--83.

\bibitem{vercelloni2017improved}
Vercelloni J, Mengersen K, Ruggeri F, Caley MJ. 2017 {Improved coral population
  estimation reveals trends at multiple scales on Australia’s Great Barrier
  Reef}.
\newblock \emph{Ecosystems} \textbf{20}, 1337--1350.

\bibitem{gonzalez2020monitoring}
Gonzalez-Rivero M, Beijbom O, Rodriguez-Ramirez A, Bryant DE, Ganase A,
  Gonzalez-Marrero Y, Herrera-Reveles A, Kennedy EV, Kim CJ, Lopez-Marcano S,
  \emph{et~al.} 2020 Monitoring of coral reefs using artificial intelligence: A
  feasible and cost-effective approach.
\newblock \emph{Remote Sensing} \textbf{12}, 489.

\bibitem{spalding2007marine}
Spalding MD, Fox HE, Allen GR, Davidson N, Ferda{\~n}a ZA, Finlayson M, Halpern
  BS, Jorge MA, Lombana A, Lourie SA, \emph{et~al.} 2007 Marine ecoregions of
  the world: a bioregionalization of coastal and shelf areas.
\newblock \emph{BioScience} \textbf{57}, 573--583.

\bibitem{Lindgren2015}
Lindgren F, Rue H. 2015 Bayesian spatial modelling with {R-INLA}.
\newblock \emph{Journal of Statistical Software} \textbf{63}, 1--25.

\bibitem{rue2009approximate}
Rue H, Martino S, Chopin N. 2009 {Approximate {B}ayesian inference for latent
  Gaussian models by using integrated nested {L}aplace approximations}.
\newblock \emph{Journal of the royal statistical society: Series b (statistical
  methodology)} \textbf{71}, 319--392.

\bibitem{souter2020status}
Souter D, Planes S, Wicquart J, Logan M, Obura D, Staub F. 2020 Status of coral
  reefs of the world: 2020.
\newblock \emph{{Global Coral Reef Monitoring Network; International Coral Reef
  Initiative, Australian Institute of Marine Science: Townsville, Australia}} .

\bibitem{moraga2019geospatial}
Moraga P. 2019 \emph{{Geospatial health data: Modeling and visualization with
  R-INLA and Shiny}}.
\newblock CRC Press.

\bibitem{vercelloni2020forecasting}
Vercelloni J, Liquet B, Kennedy EV, Gonz{\'a}lez-Rivero M, Caley MJ, Peterson
  EE, Puotinen M, Hoegh-Guldberg O, Mengersen K. 2020 Forecasting intensifying
  disturbance effects on coral reefs.
\newblock \emph{Global change biology} \textbf{26}, 2785--2797.

\bibitem{kennedy2020coral}
Kennedy EV, Vercelloni J, Neal BP, Bryant DE, Ganase A, Gartrell P, Brown K,
  Kim CJ, Hudatwi M, Hadi A, \emph{et~al.} 2020 {Coral reef community changes
  in Karimunjawa National Park, Indonesia: Assessing the efficacy of management
  in the face of local and global stressors}.
\newblock \emph{Journal of Marine Science and Engineering} \textbf{8}, 760.

\bibitem{RN1}
{Australian Institute of Health and Welfare}. 2021 {Cancer in Australia 2021}.
\newblock Report, AIHW.

\bibitem{RN2}
Leroux BG, Lei X, Breslow N. 2000 \emph{Estimation of disease rates in small
  areas: a new mixed model for spatial dependence}, pp. 135--178.
\newblock New York: Springer.

\bibitem{RN3}
{Australian Bureau of Statistics}. 2011 {Australian Statistical Geography
  Standard (ASGS): Volume 1 - Main structure and greater capital city
  statistical areas, July 2011}.
\newblock Report, ABS.

\bibitem{duncan2019development}
Duncan EW, Cramb SM, Aitken JF, Mengersen KL, Baade PD. 2019 Development of the
  {A}ustralian {C}ancer {A}tlas: spatial modelling, visualisation, and
  reporting of estimates.
\newblock \emph{International Journal of Health Geographics} \textbf{18},
  1--12.

\bibitem{duncanebook}
Duncan EW, Cramb SM, Baade P, Mengersen KL, Saunders T, Aitken JF. 2020
  \emph{Developing a Cancer Atlas using Bayesian Methods: A Practical Guide for
  Application and Interpretation}.
\newblock Cancer Council Queensland and Queensland University of Technology.
\newblock \url{https://atlas.cancer.org.au/developing-a-cancer-atlas/}

\end{thebibliography}

\end{document}